\documentclass[10pt,journal,compsoc]{IEEEtran}
\usepackage{soul}
\setuldepth{Berlin}
\usepackage{comment}
\usepackage{url}
\usepackage{algorithm}
\usepackage{algorithmic}
\usepackage{nth}
\usepackage{amssymb}
\usepackage{epsfig}
\usepackage{wrapfig}
\usepackage{enumitem}
\usepackage{subfig}
\usepackage{color}
\usepackage{multirow}

\usepackage{graphicx}

\usepackage[utf8]{inputenc}
\usepackage[english]{babel}
\usepackage{amsmath}
\usepackage{amsthm}

\ifCLASSOPTIONcompsoc
  \usepackage[nocompress]{cite}
\else
  \usepackage{cite}
\fi
\ifCLASSINFOpdf
\else
\fi
\begin{document}
%

\title{Structure-Preserving Community In A Multilayer Network: Definition, Detection, And Analysis}
\title{Efficient Approach to Compute Structure- And Semantics-Preserving Community In a Heterogeneous Multilayer Network}
\title{Computing Structure- And Semantics-Preserving Community in a Heterogeneous Multilayer Network}
\title{An Efficient Framework for Computing Structure- And Semantics-Preserving Community in a Heterogeneous Multilayer Network}
\title{Structure- And Semantics-Preserving Community Definition and Detection For Heterogeneous Multilayer Networks}
\title{Structure- And Semantics-Preserving Community Definition and Its Computation For Heterogeneous Multilayer Networks}
\title{Community Definition For Heterogeneous MLNs And Efficient Algorithms For Its Computation}
\title{A New Community Definition For Heterogeneous Multilayer Networks And An Approach For Its Efficient Computation}
\title{\huge A New Community Definition For MultiLayer Networks And A Novel Approach For Its Efficient Computation}
%
%
%
%

\author{\IEEEauthorblockN{Abhishek Santra\IEEEauthorrefmark{1},
Kanthi Sannappa Komar\IEEEauthorrefmark{2}, Sanjukta Bhowmick\IEEEauthorrefmark{3} and
Sharma Chakravarthy\IEEEauthorrefmark{4}}
\\
\IEEEauthorblockA{\IEEEauthorrefmark{1}\IEEEauthorrefmark{2}\IEEEauthorrefmark{4}IT Lab and CSE Department, University of Texas at Arlington, Arlington, Texas \\
\IEEEauthorrefmark{3}CSE Department, University of North Texas, Denton, Texas \\
Email: \IEEEauthorrefmark{1}abhishek.santra@mavs.uta.edu,
\IEEEauthorrefmark{2}kanthisannappa.komar@mavs.uta.edu,\\
\IEEEauthorrefmark{3}sanjukta.bhowmick@unt.edu,
\IEEEauthorrefmark{4}sharmac@cse.uta.edu}}

\markboth{Journal of \LaTeX\ Class Files,~Vol.~14, No.~8, August~2015}%
{Shell \MakeLowercase{\textit{et al.}}: Bare Demo of IEEEtran.cls for Computer Society Journals}
%




\IEEEtitleabstractindextext{%
\begin{abstract}
    With \textbf{M}ulti\textbf{L}ayer \textbf{N}etworks (\textbf{MLN}s) gaining popularity for modeling and analysis, it is important to have a community definition and efficient algorithms that preserve MLN characteristics. Currently, communities for MLNs, are determined by aggregating them into single graphs using type-independent or projection-based techniques followed by the application of single graph community detection  (Louvain, Infomap, ...) leading to information loss. To the best of our knowledge, we propose, for {\em the first time, a community definition for \textit{He}terogeneous \textit{MLN}s (\textit{HeMLN}s) that preserves semantics and structure}. Additionally, our definition is extensible to match diverse analysis objectives.

\indent The community definition proposed in this paper is compatible with and is an extension of the traditional definition for single graphs. We present a framework for its efficient computation using recently proposed \textit{decoupling approach}. First, we define a \textit{k-community} for \textit{k} connected HeMLN layers. Then we propose a family of algorithms for its computation using the concept of bipartite graph pairings. For broader analysis, we introduce several pairing algorithms and weight metrics for composing layer communities using participating community characteristics. This results in an extensible \textit{family of community computations}. We provide elaborate experimental results for showcasing efficiency and analysis flexibility of proposed computation using IMDb and DBLP data-sets.

    \end{abstract}


}

\maketitle

\IEEEdisplaynontitleabstractindextext

%
\IEEEpeerreviewmaketitle


%
%
%
%

\section{Motivation}
\label{sec:introduction}
\IEEEPARstart{A}{s} data sets become large and complex with multiple entity and feature types, the approaches needed for their modeling and analysis also warrant extensions and/or new alternatives to match the data set complexity, analysis flexibility, and scale. With the advent of social networks, we have already seen a surge in the use of graph-based modeling along with a renewed interest in aggregation properties, such as community and centrality (e.g., hubs), used for their analysis.
It has been shown~\cite{ArXiv/SantraKBC1-2} that data sets that may not be inherently graph-based (e.g., IMDb data set used in this paper) may also benefit from the use of graph representation for modeling (from an understanding perspective) and for performing various kinds of analysis that may be difficult or not possible using the traditional Relational Database Management System (RDBMSs) or mining approaches. 

Informally, MultiLayer Networks\footnote{The terminology used for variants of multilayer networks varies drastically in the literature and many times is not even consistent with one another. For clarification, please refer to~\cite{MultiLayerSurveyKivelaABGMP13} which provides an excellent comparison of terminology used in the literature, their differences, and usages clearly.} (or MLNs) are \textit{layers of networks} where each layer represents a simple graph and captures the semantics of one or more attributes (or features) of an entity type (as node) using an edge to represent that relationship. The layers can also be connected. If each layer of a MLN has the \texttt{same set of entities of the same type}, it is termed a Homogeneous MLN (or HoMLN.) For a HoMLN, intra-layer edges are shown explicitly and inter-layer edges are implicit (and not shown.) If \texttt{the set and types of entities are different for each layer}, then relationships of entities across layers are also shown using explicit inter-layer edges. This distinguishes a Heterogeneous MLN (or HeMLN). 

Modeling of complex data sets with multiple entity and feature types using single graphs (or even attributed graphs) has been used but is not ideal. The representation either loses information~\cite{MultiLayerSurveyKivelaABGMP13,DeDomenico201318469} for entities and features as nodes, edges, and their labels are not differentiated. Single graph representation also makes aggregate computations on subsets of entities and features difficult to analyze due to lack of corresponding algorithms. 

On the other hand, the same data sets represented as MLNs avoid information loss and are flexible in separating subsets for analysis as has been established~in \cite{Wilson:2017:CEM:3122009.3208030}, ~\cite{Boccaletti20141, MultiLayerSurveyKivelaABGMP13,BDA/SantraB17,CommSurveyKimL15, DBLP:conf/bigda/ChakravarthySK19}, 
but \textit{HeMLNs currently lack a community definition and concomitant algorithms.} This is where the contribution of this paper is directed in generalizing the established community definition for HeMLNs along with an efficient computation model using a novel approach. Among the alternatives used in the literature modularity-based community definition (which hierarchically maximizes the concentration of edges within modules (or communities) compared with random distribution of links between all nodes regardless of modules), seems to have consensus for a single graph along with several implementations that are used widely (e.g., Louvain~\cite{DBLP:Louvain}). 



For a simple graph, the current community definition preserves its structure and semantics in terms of node/edge labels and relationships. Preserving the structure of a community of a MLN (especially HeMLN) entails preserving its multilayer network structure as well as semantics of node/edge types, labels, and importantly inter-layer relationships. In other words, each HeMLN community should be a MLN in its own right. Contributions of this paper are:
 
\begin{itemize}

\item Definition of \textit{structure and semantics preserving} community for a HeMLN and an approach for its efficient computation (Section~\ref{sec:hemln-community}),
\item Identification of a composition function and formalizing the \textit{decoupling approach} for HeMLN community detection algorithms (Section~\ref{sec:community-detection}),
\item Two new bipartite pairing  algorithms for composing layers which are more appropriate for HeMLN communities (Sections~\ref{sec:alternate-bm-algos} and ~\ref{sec:mwrm-and-mwmt-algos}.) Also, identification of useful weight metrics and their relevance (Section~\ref{sec:customize-maxflow}),
\item Evaluation of our definition against the available ground truth with the widely-used modularity of detected communities (Section~\ref{sec:evaluation}), and 
\item Demonstration of using this community definition for mapping analysis objectives (Section~\ref{sec:application-and-analysis}),
\item Experimental analysis using the IMDb and DBLP data sets to establish the structure and semantics preservation (drill-down capability) of the proposed approach along with efficiency aspects  (Section~\ref{sec:experiments}.)
\end{itemize}

The paper is organized as indicated above with related work in Section~\ref{sec:related-work} and conclusions in Section~\ref{sec:conclusions}.



\section{Semantics Preservation And Efficiency}
\label{sec:structure-preservation}

Lack of a community definition for a HeMLN has resulted in various \textit{ad hoc} approaches to leverage the single graph community definition and algorithms for its detection. As a consequence, although modeled as MLNs for semantic superiority, they are reduced to a single graph for the purpose of community detection. There have been some attempts to detect (rather extract) communities on the MLN. ~\cite{Wilson:2017:CEM:3122009.3208030}  proposes multilayer extraction for \textit{Homogeneous MLNs}, such as Multilayer social network, transportation network, and collaboration network. They use the notion of vertex neighborhoods with a refinement procedure, to produce a subfamily of high-scoring vertex layer sets. ~\cite{InfoMap2014} focuses on higher-order network flows \textit{again in Homogeneous MLNs}. They extend the single graph Infomap algorithm to capture clusters for multilayer networks. All of these approaches do not preserve the structure of the HoMLNs in their results.


Our goal and focus in this paper has been pragmatic from a big data analytics perspective where analysis objectives need to be mapped to appropriate graph properties and their computation. In addition, drill down analysis of the results is critical demanding the preservation of structure as well as semantics of computed results.  The importance of these are discussed below.


\subsection{Structure and Semantics Preservation}


\begin{figure*}[h]
   \centering \includegraphics[width=\linewidth]{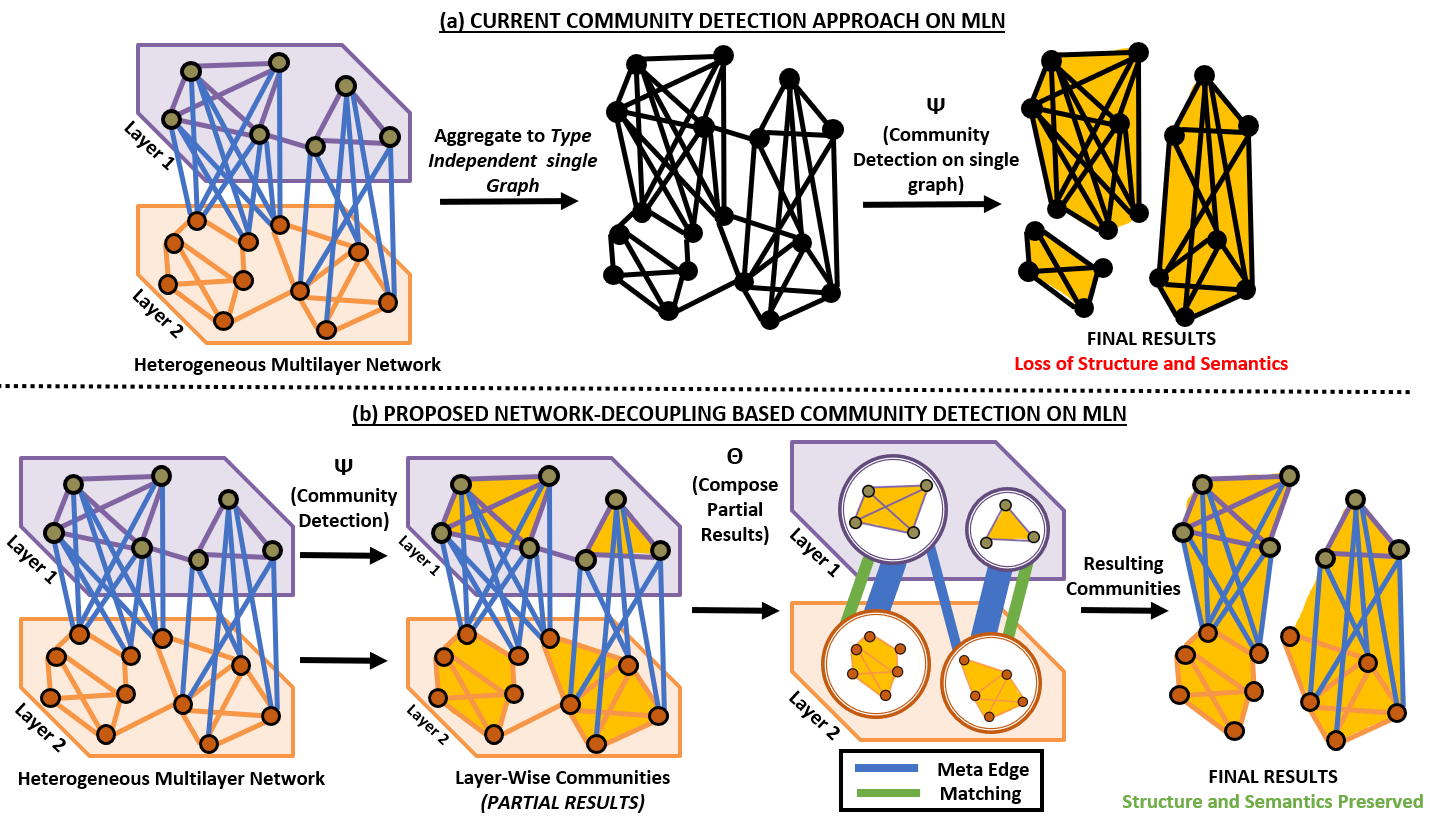}
   \caption{Traditional Lossy Approach Vs. Structure and Semantics Preserving Approach}
   \label{fig:HeMLN-comparison}
\end{figure*}

Current approaches, such as type-independent~\cite{LayerAggDomenicoNAL14} and projection-based~\cite{Berenstein2016, sun2013mining}, do not accomplish structure and semantics preservation as they aggregate (or collapse) layers into a simple graph in different ways. More importantly, aggregation approaches are likely to result in some information loss~\cite{MultiLayerSurveyKivelaABGMP13}, distortion of properties~\cite{MultiLayerSurveyKivelaABGMP13}, or hide the effect of different entity types and/or different intra- or inter-layer relationships as elaborated in~\cite{DeDomenico201318469}. Furthermore, structure and semantics preservation is critical for understanding a HeMLN community and for drill-down analysis of results.

From an analysis perspective, lack of structure and semantics makes the drill down extremely difficult (or even impossible) and hence the understanding and visualization of results. Our computation results clearly show the community structure and how easy it is to drill down to see patterns in terms of original labels and relationships.

Figure~\ref{fig:HeMLN-comparison}  illustrate the difference between the current approaches and our proposed approach.  Fig.~\ref{fig:HeMLN-comparison} a) shows type-independent aggregation\footnote{Other aggregation approaches have the same problem.} of two layers into a single  graph on which extant community detection is applied. As can be seen, \textbf{both structure as well as entity and relationship labels -- shown as colored nodes and edges -- are lost in the resulting communities.} In contrast, the Fig.~\ref{fig:HeMLN-comparison} b) shows the same layers and community detection using the definition and the decoupling approach proposed in this paper. As there is no aggregation, both structure and semantics are preserved. 


\subsection{Decoupling Approach For Efficiency}
\begin{figure}[h]
   \centering
   \includegraphics[width=\linewidth]{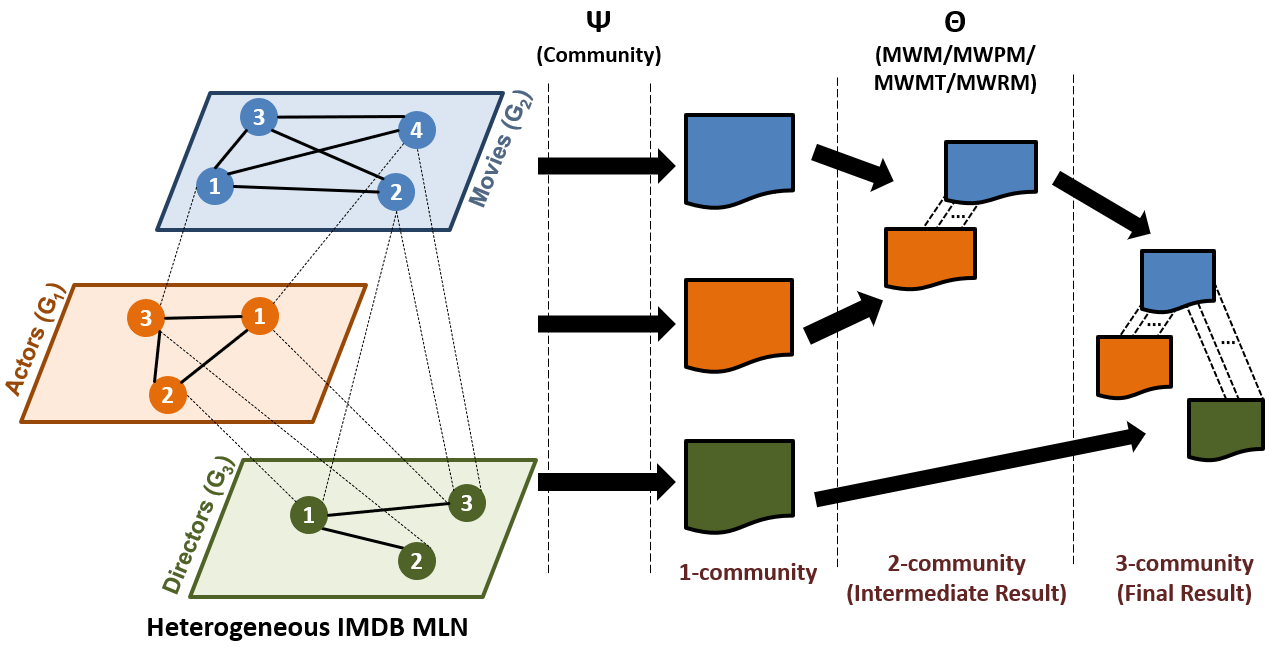}
   \caption{Decoupling Approach to Compute HeMLN 3-community Expressed as (($G_2$ $\Theta_{2,1}$ $G_1$)  $\Theta_{2,3}~G_3$); $\omega_e$\protect\footnotemark}
   \label{fig:decoupling}
\end{figure}

Decoupling approach\footnote{Although we are focusing on the decoupling approach for community detection in HeMLNs, this approach  has been shown~\cite{ICCS/SantraBC17,ICDMW/SantraBC17} to be applicable for community and centrality detection in HoMLNs.}, as proposed in this paper, is the equivalent of ``divide and conquer" for MLNs. Research on modeling a data set as a MLN \textit{and} computing on the whole MLN has not addressed efficiency issues~\cite{Wilson:2017:CEM:3122009.3208030}. Decoupling requires partitioning (derived from the MLN structure) and a way to compose partial (or intermediate) results. In this paper we identify a composition function (referred to as $\Theta$, see Figure~\ref{fig:decoupling}) that is appropriate for efficient community detection (referred to as $\Psi$, see Figure~\ref{fig:decoupling}) on MLNs. 

Fig.~\ref{fig:decoupling} shows the \textbf{proposed decoupling approach}.  Three layers and some inter-layer connections are shown. HeMLN community computation is accomplished by combining communities from two layers of a HeMLN using a composition function ($\Theta$) and is extended to $k$ layers by  composing the result with additional layers one at a time. Figure~\ref{fig:decoupling} also shows how a 3 layer HeMLN community is expressed for computation. This approach of partitioning and composing partial results is central to efficiency of computation as elaborated in Section~\ref{sec:experiments}.

\protect\footnotetext{Technically, this should be expressed as (($\Psi$($G_2$) $\Theta_{2,1}$ $\Psi$($G_1$))  $\Theta_{2,3}$ $\Psi$($G_3$).) However, we drop $\Psi$ for simplicity. In fact, $\Theta$ with its subscripts is sufficient for our purpose due to pre-defined precedence (left-to-right) of $\Theta$. We retain G for clarity of the expression. $\omega_e$ is a weight metric discussed in Section~\ref{sec:customize-maxflow}.}


\section{Related Work}
\label{sec:related-work}

As the focus of this paper is community definition and its efficient detection in HeMLNs, we present relevant work on simple graphs and MLNs. The advantages of modeling using MLNs are discussed in~\cite{Boccaletti20141,MultiLayerSurveyKivelaABGMP13,BDA/SantraB17,CommSurveyKimL15}.

\textit{Community detection} on a simple graph involves identifying groups of vertices that are more connected to each other than to other vertices in the network/graph. Most of the  work in the literature considers \textbf{single networks or simple graphs} where this objective is translated to optimizing network parameters such as modularity ~\cite{clauset2004} or conductance ~\cite{Leskovec08}. As the combinatorial optimization of community detection is NP-complete ~\cite{Brandes03}, a large number of competitive approximation algorithms have been developed (see reviews in ~\cite{Xie2013,Fortunato2009}.)
Algorithms for community detection have been developed for different types of input graphs including directed ~\cite{Yang10,Leicht08}, 
edge-weighted ~\cite{Berry2011}, and dynamic networks~\cite{porterchaos13}. 
However, to the best of our knowledge, there is no  community definition and detection that include node and edge labels, node weights as well as graphs with self-loops and multiple edges between nodes\footnote{This is in contrast to subgraph mining, search, and querying of graphs where attributed graphs are widely used.}.
Even the most popular community detection packages such as Infomap \cite{InfoMap2014} or Louvain~\cite{DBLP:Louvain}, do not accept non-simple graphs. 

Recently, community detection algorithms have been extended to \textbf{HoMLNs} (see reviews \cite{CommSurveyKimL15,CommFortunatoC09}.)  Algorithms based on matrix factorization, 
cluster expansion philosophy,  
Bayesian probabilistic models,  
regression, 
and spectral optimization of the modularity function based on the supra-adjacency representation 
have been developed.
However, all these approaches \textit{analyze a MLN either by aggregating all (or a subset of) layers of a HoMLN using Boolean and other operators or by considering the entire MLN as a whole}. Recently decoupling-based approaches for detecting communities~\cite{ICCS/SantraBC17} and centrality~\cite{ICDMW/SantraBC17} in HoMLN have been proposed, where intermediate analysis results from individual layers (obtained by existing single graph algorithms) are combined \textit{systematically in a loss-less manner} to compute communities or centrality hubs for combinations of layers.



To the best of our knowledge, there is no community definition or detection algorithm for HeMLNs. Majority of the work on analyzing HeMLN (reviewed in \cite{shi2017survey,sun2013mining}) focuses on developing meta-path based techniques for determining the similarity of objects~\cite{wang2016relsim}, classification of objects~\cite{wang2016text}, predicting the missing links~\cite{zhang2015organizational}, ranking/co-ranking~\cite{shi2016constrained} and recommendations~\cite{shi2015semantic}. 


The type-independent~\cite{LayerAggDomenicoNAL14} and projection-based~\cite{Berenstein2016} approaches used for HeMLNs \textit{neither preserve the structure nor the semantics of the community.} The type independent approach collapses  all layers into a simple graph keeping \textit{all} nodes and edges (including inter-layer edges) sans their types and labels. The same is true for the projection-based approach as well.
The presence of different sets of entities in each layer and the presence of intra-layer edges makes structure-preserving definition more challenging for HeMLNs and also warrants a novel composition technique (as proposed in this paper.) A few existing works have proposed techniques for generating clusters of entities \cite{melamed2014community}, but they have only considered the inter-layer links and not the networks themselves.

This paper fills the gap for a \textit{structure and semantics preserving HeMLN community} definition and its efficient computation.


\section{Definitions}
\label{sec:problem-statement}

A {\bf graph} $G$ is an ordered pair $(V, E)$, where $V$ is a set of vertices and $E$ is a set of edges. An edge $(u,v)$ is a 2-element subset of the set $V$. The two vertices that form an edge are said to be adjacent or neighbors. In this paper we only consider graphs that are undirected.




A {\bf multilayer network}, $MLN (G, X)$, is defined by two sets of graphs: i) The set $G = \{G_1, G_2, \ldots, G_N\}$ contains graphs of N individual layers $L = \{L_1, L_2, \ldots, L_N\}$ as defined above, where $G_i (V_i, E_i)$ is defined by a set of vertices, $V_i$ and a set of edges, $E_i$. An edge $e(v,u) \in E_i$, connects vertices $v$ and $u$, where $v,u\in V_i$ and ii) A set $X =\{X_{1,2}, X_{1,3}, \ldots, X_{N-1,N}\}$ consists of bipartite graphs. Each graph $X_{i,j} (V_i, V_j, L_{i,j})$ is defined by two sets of vertices $V_i$ and $V_j$, and a set of edges (also called links or inter-layer edges) $L_{i,j}$, such that for every link $l(a,b) \in L_{i,j}$,  $a\in V_i$ and $b \in V_j$, where $V_i$ ($V_j$) is the vertex set of graph $G_i$ ($G_j$.) 

For a HeMLN, $X$ is explicitly specified.
Without loss of generality, we assume unique numbers for nodes across layers and disjoint sets of nodes across layers\footnote{Heterogeneous MLNs can also be defined with overlapping nodes across layers (see \cite{MultiLayerSurveyKivelaABGMP13}) which is not considered  in this paper.}.
We define a \textit{k-community} to be a multilayer community where communities from $k$ distinct connected layers of a HeMLN are combined in a \textit{specified order} as shown in Figure~\ref{fig:decoupling}. Our proposed algorithm using the decoupling approach for finding HeMLN communities can be described as follows with reference to 
Figure~\ref{fig:decoupling}:
 
 {\bf(i)} First, use the function $\Psi$ (here community detection) to find communities in each of the layers individually (can also be done in parallel),
 
 {\bf(ii)} For any two chosen layers, use the partial/intermediate results from these layers and apply the composition function $\Theta$ (bipartite graph matching)  using the meta edges  (whose weight is denoted by $\omega$) between the layers  to compute the result. For HeMLN community detection, a bipartite pairing that maximizes modularity (or total weight of the meta edges) is used for $\Theta$.
 
 {\bf(iii)} The binary composition of step ii) is applied for determining a k-community for a specified order of layers.
 
Figure~\ref{fig:decoupling} illustrates the decoupling approach for specifying and computing a HeMLN 3-community from partial results. It illustrates how a set of distinct communities from a layer is used for computing a 2-community ($G_2$ $\Theta_{2,1}$ $G_1$) for 2 layers and further a 3-community (($G_2$ $\Theta_{2,1}$ $G_1$)  $\Theta_{2,3}~G_3$) for 3 layers using partial results. 1-community is the set of communities generated for a layer $L_i$ using its $G_i$ (simple graph.) We use $L_i$ and $G_i$ interchangeably in the rest of the paper.

We can now define the problem addressed in this paper.  For a given HeMLN and a set of analysis objectives, determine the appropriate triad of ~$\Psi$,  $\Theta$, and $\omega$, and a $k$-community expression for computing \textit{\textbf{each objective}}. For this paper, community is used for $\Psi$ and \textit{bipartite match algorithms for $\Theta$} for HeMLN community detection along with defining and identifying $\omega$.

\section{Community Definition for a H\MakeLowercase{e}MLN}
\label{sec:hemln-community}

We first intuitively motivate the  need for structure and semantics preserving community for a HeMLN using one of the data sets used in this paper.  

\subsection{Intuition Behind a HeMLN Community}

\noindent The IMDb data set captures movies, TV episodes, actors, directors, and other related information, such as rating, genre, etc. This is a large data set consisting of movie and TV episode data from their beginnings. This data set can be modeled and analyzed in multiple ways as well for different purposes~\cite{DaWaK/SantraSBC20}. For the IMDb data set, consider the HeMLN shown in Figure~\ref{fig:decoupling} that has the following three layers: i) \textit{Actors} layer --  connects actors who act in similar genres frequently (intra-layer edges.),  ii) \textit{Directors} layer --  connects directors who direct similar movie genres frequently, and iii) \textit{Movies} layer -- connects movies within the same rating range.  The inter-layer edges depict \textit{acts-in-a-movie, directs-a-movie} and \textit{directs-an-actor}.


Consider the analysis objective \textit{``Find dense \ul{groups} of actors and directors that have high/strong interaction/coupling with each other"} Note that, individually, the actor and director layers can only compute dense groups of actors or directors, who act in or direct similar genre, respectively. The connection (or coupling) between directors and actors only come from inter-layer edges. It is only by identifying the proper meta edges \textbf{\textit{and}} preserving the structure of both the communities in actor and director layers as well as the inter-layer edges, can we compute and drill down the answer that indicates the semantics of which actor groups are paired with which director groups. The inter-layer edges preserve the relationships of individual actors and directors as well. Preservation of structure (inter-layer edges) and semantics (labels) is critical for drilling-down to understand the results.

Clearly, multiple strong interactions can exist between groups of actors and directors (in general, among communities from different layers.) A specific co-actor group may be favorites of one or more director groups based on genre or other characteristics, and vice-versa. So, any MLN community definition needs to include these multiple couplings (unlike traditional bipartite matching which identifies only unique pairs) in a way similar to the coupling between nodes in a single layer community definition. In addition, it may also be important, from an analysis perspective, and useful to couple these groups (or communities) using different community characteristics as well.  An analysis objective may also want to use or specify different community interactions as \textit{preferences} to meet an analysis objective. As an example, one may be interested in groups (or communities) where the \textit{most important} actors  and directors (characterized in terms of their degree) interact rather than the actor community as a whole. Based on this observation, we have proposed two new bipartite matching algorithms in this paper.

Note that the community definition and detection research in the literature for homogeneous MLNs~\cite{BDA/SantraB17,ICCS/SantraBC17} are not applicable to HeMLNs as each layer has \textit{different sets and types of entities} with  \textit{inter-layer edges} between them.
It is important that this formulation of communities preserves entity and feature types as compared to other alternatives proposed in the literature.

\textit{Hence, the challenge for the definition of a HeMLN community is to not only keep it consistent with the widely-accepted community definition, but also provide alternatives to accommodate broader analysis objectives.} This, in conjunction with structure and semantics preservation, will enhance the utility of this modeling as well as analysis efficiency. In the following sections, we provide such a definition, its relationship to modularity for illustration, its efficient computation with  algorithms based on the decoupling approach, and importantly demonstrate its usage with respect to diverse analysis objectives later in the paper.

\subsection{Formal Definition of a HeMLN Community}

A \textbf{Community Bipartite Graph} or \textbf{CBG}$_{i,j}$($U_i$, $U_j$, $L'_{i,j}$)\footnote{We defined the set X of bipartite graphs between layers of HeMLN in Section~\ref{sec:problem-statement}. This is a different bipartite graph between two sets of nodes (termed meta nodes) from two distinct layers where each (meta) node corresponds to a community in each layer. A single bipartite edge (termed meta edge) is drawn between distinct meta node pairs if there are inter-layer edges between those two communities.} between graphs (layers) $G_i$ ($L_i$) and $G_j$ ($L_j$) is defined as the graph with disjoint and independent nodes $U_i$ ($U_j$) corresponding to each community from $L_i$ ($L_j$), respectively, represented as a single meta node and $L'_{i,j}$ being the set of single meta edges between the nodes of $U_i$ and $U_j$ (or bipartite graph edges.) whose weight ($\omega$) corresponds to the number (or strength) of the inter-layer edges between the corresponding nodes. 
For a layer $L_i$, a \textbf{1-community} is the set of communities identified on the graph corresponding to that layer using any of the community detection algorithms.

For two layers $L_i$ and $L_j$, and their inter-layer edges $X_{i,j}$, a \textbf{HeMLN 2-community} for $G_i$ and $G_j$ is defined as the \textbf{community bipartite graph pairs} that maximize total inter-layer edge weights (along with the overall modularity) between the two community bipartite graphs. This matching (or coupling) can be defined in multiple ways. We start with the coupling being defined as the traditional bipartite pairings that maximizes total edge weight and extend it.

A \textbf{HeMLN k-community} is the application of the above binary definition for k layers in a specified order of layers using the previously computed \textbf{(k-1)-community}..



\subsection{Need For Alternative Bipartite Match Algorithms}
\label{sec:alternate-bm-algos}

Traditional bipartite graph matching with edge weights and different size node sets compute pairings (or matchings) that produce maximum weight (termed MWM or Maximum Weight Match~\cite{kuhn1955hungarian, edmonds1965maximum}) or that produce maximum number of pairings with maximum weight (MWPM or Maximum Weight Perfect Match~\cite{hopcroft1973n,gabow2017weighted}). A constraint used in all traditional bipartite matches is that the resulting matches/pairs are unique. That is, \textit{a single node of one set will be matched with at most one node of another set}. This was appropriate for applications such as job assignment, hiring, and pairing.

However, for a HeMLN community definition that maximizes the coupling between bipartite graphs, the above can be used directly \textit{if one is interested in unique pairings from an analysis perspective}. However, for many analysis using communities, this is not suited. We need pairings (couplings) without the above restriction as one-to-many pairings (from either side) makes sense from an analysis perspective. For the analysis of the IMDb data set mentioned earlier, there is no reason to restrict an actor group to pair with only one director group if another coupling is equally strong or an alternate coupling produces a higher total weight. Hence, we need to: i) relax the unique match restriction to increase total edge weight for the \textit{same number}\footnote{If one wants to merely maximize the edge weights under relaxed uniqueness pairing, one could just choose top k edges (weight-wise) which will give k pairings. However, this will be maximal and will not have any global properties. Hence, we have chosen to extend the MWM for this relaxation and include ties as they provide a global maximum to start with.} of pairings and ii) deal with (or include) ties of edge weights incident on the same nodes of unique pairings (instead of choosing one randomly!.) These two will maximize $\sum w(e)$ of the CBG across all edges and minimize the number of such pairs. These relaxations make sense semantically as well as a community may have a stronger coupling with multiple communities. These are global maximums as they are derived from the traditional pairings of MWM and MWPM.  We believe that MWM (and the variants we are proposing) comes closest to modularity semantics for HeMLN communities by maximizing the inter-layer connectivity for the communities in contrast to inter-layer connectivity of other communities. 


 \begin{figure}[ht]
   \centering
   \includegraphics[width=\linewidth]{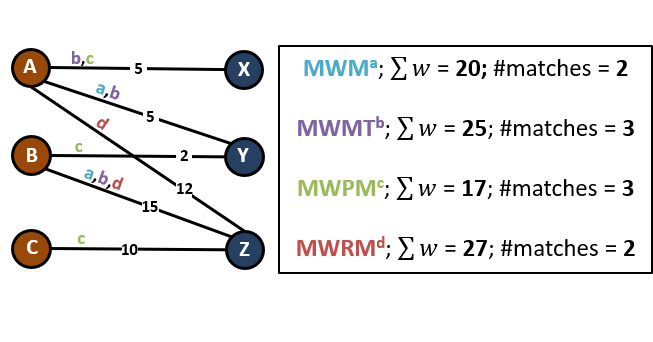}
\vspace{-15pt}
   \caption{{Illustration of Traditional and Relaxed Pairings on a weighted bipartite graph}}
   \label{fig:sequence-and-pairalgo}
\end{figure}

Figure~\ref{fig:sequence-and-pairalgo} provides an example of a bipartite graph to illustrate
the above discussion. MWM(Maximum Weight Matching); MWMT (MWM with Ties); MWPM(Maximum Weight Perfect Match); MWRM(Maximum Weight with Relaxed Matching).
Relaxing the unique pair constraint can increase the maximum weight if alternative pairings exist and they are not unique. In addition, the presence of ties results in additional pairings to maximize modularity under our relaxation. Our matchings are termed MWMT (Maximum Weight Matching with Ties) and MWRM (Maximum Weight with Relaxed Matching where the unique pairing constraint is relaxed). All of the above are commutative and non-associative. Other possibilities also exist\footnote{Instead of pairing globally for a metric, it is possible to pair or couple locally from a given side by pairing each node using the maximum weight possible for that node. This corresponds to the \textit{maximal} versions and gives a local maximum. Note that this can be done from either side with different results and hence not commutative. it is also not associative. This may be relevant to analysis objective mapping. Refer to~\cite{Arxiv/SantraKBC19,msThesis/Komar19} for details.}.

Graph communities can be disjoint or overlapping. Our definition works for both. If we assume disjoint 1-community from each layer, the traditional matching algorithms (MWM and MWPM) give disjoint HeMLN communities by definition. On the other hand, even if the layer communities are disjoint, the relaxed bipartite matches (MWRM and MWMT) can give overlapping HeMLN communities. If the initial layer communities are overlapping, both traditional and relaxed pairings give overlapping communities. 


Although, in the above definition, we have chosen the weight of the meta edge of the bipartite graph from a modularity perspective, this weight can include (or reflect) other participating community characteristics to accommodate a family of HeMLN community definitions as elaborated in Section~\ref{sec:customize-maxflow}. 

\textit{Most importantly, unlike current alternatives in the literature for community of a MLN, there is no need for aggregating or collapsing a MLN into a single graph in our definition and computation, thereby avoiding information loss or distortion or hiding the effect of different entity types or relationships. The representation of a HeMLN community preserves the MLN structure along with semantics (node and edge labels, both intra and inter.)}


\subsection{HeMLN k-Community Computation}

This section outlines the computation of the HeMLN community definition given above for an arbitrary HeMLN. Although the above definition is commutative, it is not associative. Hence, different HeMLN communities can be obtained depending on the order used for its computation\footnote{This corresponds to the different communities used as bipartite graph nodes -- somewhat similar to different nodes as starting points for community detection of a simple graph.}. Due to space constraints, we are not addressing the order issues that may exhibit certain desirable properties in this paper. The order of community computation is derived mainly from analysis objectives and is mapped to community composition expressions.

\begin{table}[h!t]
    \renewcommand{\arraystretch}{1.3}
    \centering
        \begin{tabular}{|c|p{5.6cm}|}
            \hline
            $G_i (V_i, E_i)$ & Simple Graph for layer \textit{i} or $L_i$\\
            \hline
            $X_{i,j} (V_i,V_j,L_{i,j})$ & Bipartite graph of layers \textit{i} and \textit{j} \\
            \hline
            $MLN (G, X)$ & Multilayer Network of layer graphs (set \textit{G}) and Bipartite graphs (set \textit{X})\\
            \hline
            $\Psi$ & Analysis function for $G_i$ (community)\\
            \hline
            $\Theta_{i,j}$ & Composition function for $G_i$ and $G_j$ (weighted bipartite Matching algorithms) \\
            \hline
            $CBG_{i,j}$ & Community Bipartite Graph for $G_i$ and $G_j$ \\
            \hline
            $U_i$ & Meta nodes of layer \textit{i} from 1-community \\
            \hline
            $L'_{i,j}$ & Meta edges between $U_i$ and $U_j$ \\
            \hline
            $c_i^{m}$ & $m^{th}$ community of $G_i$ \\
            \hline
            $v_i^{c^{m}}$, $e_i^{c^{m}}$ & Vertices and Edges in community $c_i^{m}$\\
            \hline
            $H_i^m$ & Hubs in $c_i^{m}$ \\
            \hline
            $H_{i,j}^{m,n}$ & Hubs in $c_i^{m}$ connected to $c_j^{n}$\\
            \hline
            $x_{i,j}$ & \{Expanded(meta edge $<c_i^{m}$, ~$c_j^{n}>$)\}\\
            \hline
            $0$ and $\phi$ & null community id and empty $x_{i,j}$\\
            \hline
            $\omega_e$, $\omega_d$, $\omega_h$ & Weight metrics for meta edges (see Section~\ref{sec:customize-maxflow})\\
            \hline

        \end{tabular}
            \caption{Notations used in this paper}
    \label{table:notations}
\end{table}

\label{sec:definition}



Table~\ref{table:notations} lists all notations used in the paper and their meaning for quick reference. 

A \textbf{\textit{1-community}} for a given layer is computed using any community detection algorithm. The \textbf{community bipartite graph}
\textbf{\textit{CBG$_{i,j}$($U_i$, $U_j$, $L'_{i,j}$)}} is computed using $X_{i, j}$ once the 1-community for layers $L_i$ and $L_j$ are computed.

A \textbf{\textit{2-community}}, corresponding to layers $L_i$ and $L_j$, is computed on the community bipartite graph \textit{CBG$_{i,j}$($U_i$, $U_j$, $L'_{i,j}$)}. A 2-community is a set of tuples each with a pair of elements $<c_i^m, c_j^n>$, where $c_i^m \in U_i$ and $c_j^n \in U_j$, that satisfy \textit{one of the weighted bipartite matching algorithms discussed} (composition function $\Theta$ defined in Section~\ref{sec:problem-statement}) for the bipartite graph of $U_i$ and $U_j$, \textbf{along with} the set of inter-layer edges between them (denoted $x_{i,j}$.) The pairing is done using the specified pairing alternative (one of MWM, MWPM, MWRM, or MWMT) to obtain pairs of communities and their inter-layer links for $L_i$ and $L_j$. Note that it is possible that several of the matching algorithms may give the same result depending on the bipartite graph characteristics.

A \textbf{\textit{k-community}}\footnote{k represents the number of layers used for computing the community, not the number of compositions. 
A k-community corresponds to a connected subgraph of k layers. Our definition assumes left-to-right precedence for the composition function $\Theta$. It is possible to define a k-community with explicit precedence specification for $\Theta$.
Also, other definitions are possible that may be order agnostic.}  
for \textit{k} layers of a HeMLN is computed by applying the \textit{2-community} computation repeatedly as per the given expression that includes order.

We start with the 2-community of the first two layers in the expression -- termed a t-community. For each new binary computation step, there are two cases for the 2-community computation under consideration: i) the $U_i$  is from a layer $G_i$ \textit{already in the t-community} and the $U_j$  is from a \textit{new layer $G_j$}. This bipartite graph match is said to \textbf{extend} a t-community (t $<$ k) to a \textit{(t+1)-community},
or ii) \textbf{both} $U_i$ ($U_j$)  from layers $G_i$ ($G_J$) are \textit{already in the t-community}.
This bipartite graph match is said to \textbf{update} a t-community (t $<$ k), \textbf{not extend} it.


The possibility that both layers are not in the t-community is the initial step of 2-community computation. The possibility that one of them is not in the t-community is not possible for a k-community, by definition, as all the layers in the expression are connected and the expression is computed from left to right.


For both cases i) and ii) above, two outcomes are possible. Either a meta node from $U_i$: a) matches one or more meta nodes in $U_j$ resulting in one or many \textbf{consistent match}, or b) does not match a meta node in $U_j$ resulting in a \textbf{no match}. However, for case ii) above, a third possibility exists which can be characterized as  c) matches a node in $U_j$ that is not consistent with a previous match and is termed an \textbf{inconsistent match}. Since both communities have already been matched, a previous consistent match exists. If the current match is not the same, then it is an \textbf{inconsistent} match. Note that each of the relaxed pairings is a separate HeMLN k-community.



Structure preservation is accomplished by retaining, for each tuple of t-community, either a matching community id (or 0 if no match) and 
$x_{i,j}$ (or $\phi$ for empty set) representing inter-layer edges corresponding to the meta edge between the meta nodes (termed \textbf{expanded(meta edge)}). The \textit{extend} and \textit{update} carried out for each of the outcomes on the representation is listed in Table~\ref{table:algo-refer}. Note that due to multiple pairing of nodes during any composition, the number of tuples (or t-communities) may increase. Copy \& update is used to deal with multiple pairings.




\begin{table}[h!t]
\centering
    \begin{tabular}{|p{3.1cm}|p{5.0cm}|}
            \hline
                \textbf{($G_{left}$, $G_{right}$) outcome} & \textbf{Effect on tuple \textit{t} } \\
            \hline
            \hline
            \multicolumn{2}{|c|}{\texttt{\textcolor{blue}{case (i) - one processed and one new layer}}} \\
            \hline
            a) \texttt{\textcolor{blue}{consistent match}} & \textbf{Extend (Copy \& Extend)} \textit{t} with paired community id \textbf{and} $x_{i,j}$\\
            \hline
            b) \texttt{\textcolor{blue}{no match}} & \textbf{Extend (Copy \& Extend)} \textit{t} with 0 and $\phi$\\
            \hline
            \hline
            \multicolumn{2}{|c|}{\texttt{\textcolor{blue}{case (ii) - both are processed layers}}} \\
            \hline
            a) \texttt{\textcolor{blue}{consistent match}} & \textbf{Update (Copy \& Update)} \textit{t} \textbf{only with} $x$\\
            \hline
            b) \texttt{\textcolor{blue}{no match}} & \textbf{Update (Copy \& Update)} \textit{t} \textbf{only with} $\phi$\\
            \hline
            c) \texttt{\textcolor{blue}{inconsistent match}} & \textbf{Update (Copy \& Update)} \textit{t} \textbf{only with} $\phi$\\
            \hline

        \end{tabular}
            \caption{Cases and outcomes for MWxx (Extend and Update for MWPM/MWM; copy \& extend/update or update for MWRM/MWMT) used in Algorithm \ref{alg:k-community}}
    \label{table:algo-refer}
\end{table}


\subsection{Characteristics of k-community}

A HeMLN can be viewed as a simple graph (termed HeMLN-graph) with each layer of a HeMLN being a node and the \textit{presence} of inter-layer edges between layers denoted by a single edge between corresponding layer nodes. Then, a k-community can be specified over any connected subgraph of the HeMLN-graph. Case i) above corresponds to a k-community of an \textit{acyclic} subgraph of HeMLN-graph and case ii) to a k-community of a  \textit{cyclic} subgraph. For both, the number of compositions will be determined by the number of edges in the connected subgraph and can be more than the number of layers (or nodes).  Also, for both cases, MWxx algorithm results in one of the 3 outcomes: a \textit{consistent match}, \textit{no match}, or an \textit{inconsistent match (only for case (ii))} as shown in Table~\ref{table:algo-refer}.

The above definition when applied to a specification 
generates \textit{progressively strong coupling of communities between layers using specified MWxx pairing}. \textit{Thus, our definition of a k-community is characterized by dense connectivity within the layer (community definition) and semantically strong coupling across layers using one of MWxx.} Hence, we believe, that this definition of k-community matches the original intuition of a community. By refining the pairing used and the edge weight based on participating community characteristics, it supports a family of community definitions that can be customized.




\subsubsection{\bf Space of Analysis Alternatives}

Given a HeMLN with k layers and at least (k-1) inter-layer edges, the number of possible k-community (or analysis space) is quite large. For a HeMLN-graph, the number of potential k-community is a function of the number of unique connected subgraphs of different sizes and the number of possible orderings for each such connected subgraph. With the inclusion of $m$ bipartite pairing choices and $n$ weight metrics (see Section~\ref{sec:customize-maxflow}), it gets even larger. It is important to understand that each subgraph of a given size (equal to the number of edges in the connected subgraph) along with the ordering represents a \textit{different} analysis of the data set and provides a different perspective thereby supporting a large space of analysis alternatives. 

The composition function $\Theta$ defined above (one of MWM, MWPM, MWRM, MWMT) is commutative and not associative. Hence, for each k-community, the \textit{order in which a k-community} is defined has a bearing on the result (semantics) obtained. In fact, the ordering is important as it differentiates one analysis from the other even for the same set of layers  and inter-layer connections as elaborated in Section~\ref{sec:application-and-analysis}. 



\subsubsection{\bf Number of k-communities} 

For one-to-many pairing of the nodes of the bipartite graph (relaxed pairings), the theoretical upper bound is the sum of the cross product of communities in each pairing. In practice, it is likely to be significantly less (as can also be seen from the experiments.)  Each tuple generated corresponds to a k-community at the end of computation. We can  say that the number of pairings produced for MWPM is at least as many as that those produced for MWM. Since the relaxation is done on the MWM result, the number of pairings is bound by the MWM for each application of the composition function. However, the average case is likely to be much closer to the number of consistent matches (or pairs) obtained for the first 2-community. Weighted bipartite coupling in every iterative step is dictated by the number of nodes and edges for that composition. For MWM and MWPM, the upper bound on the number of pairings at each step is determined by the number of nodes in min($|U_i|$, $|U_j|$).


Also, each element of a k-community  can be total or partial. A \textbf{partial k-community element} has \textit{\textbf{at least one} $\phi$ or 0} as part of the tuple. Otherwise, it is a \textbf{total k-community element}. There may be more than one \textit{no match} in a tuple as well. Any k-community that is \textbf{total} reflects a stronger coupling as it includes all inter-layer edges for those communities (as is the case of M-A-D-M in Figure~\ref{fig:A5-IMDB-MADM-we-MWMT} in Section~\ref{sec:experiments}.) A \textbf{partial} k-community  element, on the other hand, for both acyclic and cyclic cases indicates strong coupling only among \textit{the consistent match layers}.


\noindent\subsubsection{\bf Importance of Weights} 


\label{sec:new-pairing-algo}
For traditional weighted bipartite pairing, maximum weighted matching (MWM) or maximum weighted perfect matching (MWPM) algorithms (e.g.,  ~\cite{edmonds1965maximum}) are used mainly because each node of a bipartite graph is a simple node.
In contrast,  each node of our bipartite graph is a meta node and the bipartite edge is also a meta edge. Each meta node, in our case, is a community representing a group of entities with its own characteristics (connectivity, degree, etc.) Each meta edge needs to, at the least, capture the number of edges in that meta edge (i.e., inter-layer edges.) The number of edges between the meta nodes is one of the proposed edge weights ($\omega_e$) which corresponds to the traditional intuition behind a community. 

Since edge weights play a significant role in the matching and is also used as a mechanism for determining the strength of the coupling of communities across layers, edge weights are used as a vehicle to include participating community characteristics. In addition to $\omega_e$, it is possible to bring in participating community characteristics to capture additional aspects for coupling. 
We discuss a number of alternatives for weights (termed weight metrics $\omega$) in Section~\ref{sec:customize-maxflow}, derived from real-world scenarios. 



\subsection{\bf Evaluation of Proposed Community Definition} 
\label{sec:evaluation}

Ideally one would evaluate a new community definition by comparing the result with existing definitions. Since we do not have a community definition for a HeMLN, the \textbf{closest ground truth} is the type-independent aggregation of a heterogeneous multilayer network into a single network (as shown in Figure~\ref{fig:HeMLN-comparison} a). Hence, we compare our results with this. Also, modularity is a widely accepted metric to measure the strength of division of a network into communities~\cite{newman2006modularity}. So, we use modularity for comparing our HeMLN communities (shown in Figure~\ref{fig:HeMLN-comparison} b) with the type-independent communities obtained. We have computed modularity for different weight options as well as different matching algorithms to indicate how coupling strength changes with weights and matching algorithms. Below, we compare the pairings for the default $\omega_e$ weight metric.
For evaluation purpose, we use the HeMLNs, as described in Section~\ref{sec:application-and-analysis} and whose layer details are shown in Table ~\ref{table:DBLPHeMLNStats} and \ref{table:IMDbHeMLNStats} of section~\ref{sec:experiments}. For IMDb, we have used the Actor and Director layers with their inter-layer edges. For DBLP, we have used the Author and Paper layers with their inter-layer edges. 

\begin{table}[h!t]
\renewcommand{\arraystretch}{1}
\centering
    \begin{tabular}{|c|p{1.1cm}|p{1.1cm}|p{1.1cm}|p{1.1cm}|}
        \hline
     Type-Independent & MWM & MWMT & MWPM & MWRM \\
        \hline
        
        0.777 & 0.643(83) & 0.643(220) & \textbf{0.698(95)} & 0.603(83)
        \\
        \hline
    \end{tabular}
\caption{Modularity (\# of Matches) for IMDb with A $\Theta$ D; $\omega_e$}
\label{table:modularity}
    
\end{table}

For DBLP, the modularity value for our HeMLN community (Au $\Theta$ P;  $\omega_e$) obtained with each pairing algorithm is  \textit{equal to the modularity value for the type-independent community} (0.69). Typically, modularity value greater than 0.5 is considered a \textit{good community}. Good  HeMLN communities are also obtained for IMDb (using, A $\Theta$ D; $\omega_e$). However, the tuples/matched pairs (shown in parenthesis) vary slightly with the chosen algorithm due to which the structure of the communities change leading to \textit{different} modularity values. MWPM generates the best modularity as compared to the ground truth. It was observed that the Actor and Director communities that were paired  by MWPM had a dense intra-edge connectivity and many actor-director pairs participated in the interaction, thus resulting in a high modularity. However, in case of the type-independent communities the actor and director node types get mixed up and smaller denser communities are produced leading to a higher modularity as compared to the HeMLN community, where the node types are kept separate.  
\section{H\MakeLowercase{e}MLN \MakeLowercase{k}-Community Algorithm}
\label{sec:community-detection}


In this section, we first present a specification of a k-community and elaborate on a structure preserving representation for the result. Then we present a community detection algorithm using any of the bipartite algorithms along with the proposed bipartite matching algorithms. 

\subsection{k-community Representation}
\label{sec:rep}

Linearization of a HeMLN structure is done using an order of specification which is also used for computation. Although a k-community need to be specified as an expression involving $\Psi$ and $\Theta$, as indicated earlier, we drop $\Psi$ for clarity. For the layers shown in Figure~\ref{fig:decoupling}, an example 3-community  specification is (($G_1$ $\Theta_{1,2}$ $G_2$)  $\Theta_{2,3}$ $G_3)$. We can drop the parentheses as the precedence of $\Theta$ is assumed. However, we need the subscripts for $\Theta$ to disambiguate a k-community specification when a composition is done on the layers already used. 
A 3-community involving a cycle (when an expression corresponds to a HeMLN subgraph with a cycle) can be specified as $G_1$ $\Theta_{1,2}$ $G_2$  $\Theta_{2,3}$ $G_3 ~\Theta_{3,1}~ G_1$.  

A k-community is represented as a set of tuples. Each tuple represents a distinct element of a k-community and includes an ordering of k community ids as items  (a path, if you will, connecting community ids from different layers) and at least (k-1)  expanded(meta edge) (i.e., $x_{i,j}$) elements. This representation completely preserves the MLN structure along with semantics (labels) to reconstruct a HeMLN for any k-community. It is possible that there are multiple paths originating from the communities in the first layer of the expression due to relaxed pairings. That is, a community in a layer can participate in more than one k-community tuple. All these paths need not remain total as the k-community computations progresses\footnote{This is in contrast to the traditional pairing algorithms where a community can participate in \textit{only one path of a k-community}.}. In summary, each k-community is a tuple with 2 distinct components. The first component is a comma-separated sequence of \textit{k community ids (as items)} from a layer. The second component is  also a comma-separated sequence of \textit{at least (k-1)} $x_{i,j}$ (with each x having a different pair of subscripts.) Communities for $x$ are uniquely identifiable from the subscripts.
It is exactly (k-1) if the k-community is for an acyclic connected graph and more depending upon the number of edges in cyclic subgraph. For the above example,  it is 3 as one cyclic edge is included. 
To generalize, an element of \textit{\textbf{k-community}} for an arbitrary specification 
\begin{center} $G_{n1}$ $\Theta_{n1,n2}$ $G_{n2}$ $\Theta_{n2,n3}$ $G_{n3}$ ... $\Theta_{ni,nk}$ $G_{k}$ 
\end{center} 
will be represented as 
\begin{center} $<$ $c_{n1}^{m1}$, $c_{n2}^{m2}$, ..., $c_{nk}^{mk}$ \textbf{;} $x_{n1,n2}$, $x_{n2,n3}$, ..., $x_{ni,nk}$ $>$, where some c's may be 0 and some x's may be $\phi$. 
\end{center}

\subsection{Proposed Representation Benefits}

\begin{enumerate}
    \item Each element (a k-community) preserves the structure of all 1-communities that are part of a k-community as well as the inter-layer edges explicitly.
    \item The number of elements in a k-community corresponds to the number of distinct paths (for both traditional and relaxed matchings.) Compared to traditional matching (MWPM and MWM), relaxed matching (MWRM and MWMT) is likely to produce at least the same or more k-community elements.
    \item Each element of a k-community can be further analyzed individually (or even visualized) as the tuple contains all the information to reconstruct the HeMLN and drill down for details.
    \item Total and partial elements of k-community provide important information about the result characteristics.  A partial community shows a weak coupling of the complete community whereas a total element indicates strong coupling. 
    \item The resulting set can be ranked in several ways based on HeMLN community characteristics. For example, they can be ranked based on community size or density (or any other feature) as well as significance of the layer.
    
\end{enumerate}

\subsection{MWRM and MWMT Algorthms}
\label{sec:mwrm-and-mwmt-algos}

\begin{algorithm}[h]
\caption{MWRM and MWMT Algorithms}
\label{alg:mwrm-and-mwmt}
\begin{algorithmic}[1]
\REQUIRE- \\
   \textbf{INPUT:}  Community bipartite graph (CBG) \\
   
   \textbf{OUTPUT:} edge list ($O_{el}$) for MWRM or MWMT \\

    \STATE \textbf{Initialize:} $I_{el}$ $\gets$ MWM edges of CBG\\ 
        $M_{el}$ $\gets$ Meta edges of the input bipartite graph\\
               Set $O_{el}$ to $I_{el}$; 
               \textbf{Sort} $I_{el}$ on edge weights\\
                $I_{e}$ $\gets$ edge from $I_{el}$ with the \textbf{lowest weight}\\
                
    \WHILE {$I_{e}$ not NULL}
    
       \IF {\textbf{MWRM}}
       \FOR{\textbf{each} $M_e$ $\in$ $M_{el}$ that is incident on $I_e$}
       \IF {weight($M_e$) $>$ weight($I_e$) and $M_e$ $\notin$ $O_{el}$ }
       \STATE replace $I_e$ in $O_{el}$ with $M_e$
       \ENDIF
       \ENDFOR
       \STATE $I_e$ $\gets$ next \textbf{lowest weight} edge from $I_{el}$ or NULL \\
       \ENDIF
        \ENDWHILE
\end{algorithmic}
\end{algorithm}

Algorithm~\ref{alg:mwrm-and-mwmt} shows the computation of MWRM pairing. Line 1 gets the edge list from MWM algorithm of ~\cite{edmonds1965maximum} and sorts the edges. The while loop starting in line 2 goes through this edge list from lowest weight and replaces it with a higher value edge if it has not been already chosen. This is done for all the edges in the MWM edge list. The additional complexity involves sorting MWM paired edges and only inspecting the number of edges incident on the nodes that are paired by MWM. The algorithm for MWMT is very similar except that it adds (instead of replacing) in line 6 edges that are ties.  There is no need to sort but only inspect the number of edges incident on the nodes that are paired by MWM. Both of these are substantially less than the number of inter-layer edges. Based on the characteristics of the matching algorithms, we can assert the following for the total weights, $MWPM <= MWM <= MWRM$ and $MWM <= MWMT$. Moreover, MWM  will generate the minimum number of pairs. MWM and MWRM will give same number of pairs.



\subsection{k-community Detection Algorithm}
\label{sec:flow}

\begin{algorithm}[h]
\caption{HeMLN k-community Detection Algorithm}
\label{alg:k-community}
\begin{algorithmic}[1]
\REQUIRE- \\
   \textbf{INPUT:}  HeMLN, ($G_{n1}$ $\Theta_{n1,n2}$ $G_{n2}$ ... $\Theta_{ni,nk}$ $G_{nk}$), MWM/MWPM/MWRM/MWMT), a weight metric ($\omega$). \\
   
   \textbf{OUTPUT:} Set of distinct HeMLN k-community tuples\\
    \STATE \textbf{Initialize:} k=2, $U_i$ = $\phi$, $U_j$ = $\phi$,  result$'$ = $\emptyset$\\  
    \textit{result} $\gets$ MWxx($G_{n1}$,$G_{n2}$, HeMLN, $\omega$) \\
    \textit{left}, \textit{right} $\gets$ left and right subscripts of \textbf{second} $\Theta$ \\

    \WHILE{\textit{left} $\neq$ null \&\& \textit{right} $\neq$ null}
        \STATE $U_i$ $\gets$ subset of 1-community($G_{left}, result$) \\
        \STATE $U_j$ $\gets$ subset of     1-community($G_{right}, result$) \\ 
    
        \STATE \textit{MP} $\gets$ MWxx($U_i$, $U_j$, HeMLN, $\omega$) \\
        \texttt{//a set of comm pairs $<c_{left}^x$,$c_{right}^y>$} \\        
        \FOR{ \textbf{each} tuple \textit{t} $\in$ \textit{result} } \STATE kflag = false
            \IF {\textit{both $c_{left}^x ~and~ c_{right}^y$} are part of \textit{t} and $\in$ MP \texttt{\textcolor{blue}{[case ii (already processed layers): consistent match]}}}
                \STATE Update \textit{a copy of t} with  ($x_{left, ~right}$) and append to result$'$
            \ELSIF {$c_{left}^x$ is  part of \textit{t} and $\in$ MP and $G_{right}$ layer has been processed \texttt{\textcolor{blue}{[case ii (processed layer): no and inconsistent match]}} }
                \STATE Update \textit{a copy of t} with $\phi$ and append to result$'$\\
            \ELSIF {$c_{left}^x$ is part of \textit{t} and \textcolor{red}{for each} $c_{left}^x ~\in~ MP$  \texttt{\textcolor{blue}{[case i (new layer): consistent match]}}}
                \STATE copy and Extend \textit{t} with paired $c_{right}^y$ $\in$ MP and $x_{left, ~right}$ and append to result$'$;  kflag = true \\
            \ELSIF {$c_{left}^x$ is part of \textit{t} and $\notin$ MP \texttt{\textcolor{blue}{[case i (new layer): no match]}}}
                \STATE copy and Extend \textit{t} with 0 (community id) and $\phi$ and append to result$'$; kflag = true \\
            \ENDIF
        \ENDFOR \\
        \textit{left}, \textit{right} = next left, right subscripts of $\Theta$ or null\\
        if kflag k = k + 1; result = result$'$; result$'$ = $\emptyset$
        \ENDWHILE
     \end{algorithmic}
\end{algorithm}

Algorithm~\ref{alg:k-community}  accepts a linearized specification of a k-community expression and computes the result as described earlier. The input is an \textit{ordering of layers}, \textit{composition functions indicating the community bipartite graphs algorithms to be used} and the type of weight to be used. 
The output is a \textit{set} whose \textit{elements are tuples corresponding to  distinct, single HeMLN k-community} as described earlier. The size (i.e., number of tuples) of this set is determined by the pairs obtained during computation. The layers for any 2-community bipartite graph composition are identifiable from the input specification. 

The bipartite graph for the first 2-community and for each application of $\Theta$ is constructed for the participating layers (either one is new or both are from the t-community for some t) and specified MWxx algorithm is applied. The result obtained is used to either extend or update (or copy \& extend or update)  the tuples of the t-community depending on the algorithm used. All cases are described in Table~\ref{table:algo-refer}.

The algorithm iterates \textbf{(lines 2 to 18)} until there are no more compositions to be applied. The number of 2-community computations is equal to the number of $\Theta$ in the input (corresponds to the number of inter-layer connections in the expression.) For each layer, we assume that its 1-community has been computed. 

Line 5 computes the $k^{th}$ composition. \textbf{Lines 6 to 17} apply the results of the specified MWxx algorithm  (\textbf{line 5})
to generate tuples of the $k^{th}$ composition  using the Table~\ref{table:algo-refer}. Care is taken in the composition to make sure either the tuple is updated or extended by keeping a flag and checking it after  \textbf{line 17}. The order of checking inside the for loop (\textbf{lines 6 to 17}) is important to generate the correct k-community tuples.

Note that the k-community size \textbf{is incremented} only when a \textit{new layer is composed (case i).} For case ii) (cyclic k-community) \textbf{k is not incremented} when \textit{both layers are part of the t-community}. When the algorithm terminates, we will have the set of tuples each corresponding to a single, distinct  k-community for the given specification.



\subsection{A Note on General k-community}

Although this paper only discusses the k-community whose computation order is specified, we want to emphasize that the proposed approach -- both specification and computation -- can be easily extended and generalized to define any arbitrary computation order for a  k-community. For example, by indicating  precedence for applying 2-community bipartite match, an arbitrary k-community expression can be easily defined and computed by the above algorithm with minor changes (essentially a postfix conversion of the expression). For example, the k-community specification $(((G_3$ $\Theta_{3,2}$ $G_2$)  $\Theta_{2,1}$ $G_1) ~\Theta_{2,3}~ G_3)$  could be specified alternatively as a \textit{k-community}  $((G_3$ $\Theta_{3,2}$ $G_2$)  $\Theta_{3,2}$ ($G_1 ~\Theta_{1,2}~ G_2))$ which has a totally different semantics from an analysis perspective. 

Some of these variants may also be amenable to additional parallel processing clearly validating the efficiency aspect of the decoupling approach proposed in this paper. For the above expression the results computed in parallel on $(G_3$ $\Theta_{3,2}$ $G_2$)  and  ($G_1 ~\Theta_{1,2}~ G_2)$ can be composed using $\Theta_{3,2}$. We do not discuss this further in this paper due to space constraints.



\section{Customizing The Bipartite Graph}
\label{sec:customize-maxflow}

Algorithm~\ref{alg:k-community} in Section~\ref{sec:community-detection} uses any of the specified bipartite graph match algorithms with a given weight metric. As we indicated earlier, there is an important difference between simple and \textbf{meta nodes/edges} that represent a \textbf{community of nodes along with their edges}. Without including the characteristics of meta nodes and edges for the match, we cannot argue that the pairing obtained represents analysis based on  participating community characteristics. Hence, it is important to identify how qualitative community characteristics can be mapped quantitatively to a weight metric (that is, weight of the meta edge in a community bipartite graph) to influence the bipartite matching. Below, we propose three weight metrics and their intuition.
Number of inter-community edges as weight ($\omega_e$) can be used as default.


\subsection{Number of Inter-Community Edges ($\omega_e$)}
\label{sec:m1}

This metric uses actual number of inter-community edges of participating communities as weight for meta-edges. The intuition behind this metric is \textit{maximum connectivity} (size of the community is to some extent factored into it) without including other community characteristics. This weight connotes \textit{maximum interaction between two communities}. This weight  also corresponds to the traditional community definition.

For every meta edge $(u_i^m, u_k^n)$ $\in$ $L'_{i,k}$, where $u_i^m$ and $u_k^n$ are the meta nodes corresponding to communities $c_i^m$ and $c_k^n$, respectively, in the community bipartite graph, the weight, 

\textit{\noindent $\omega_e$($u_i^m$, $u_k^n$) =  $\frac{|x_{i,k}|}{max(\omega_e)}$}, ~where \\
\textit{\noindent $x_{i,k}$  = $\bigcup~~ \{(a, b): a \in v_i^{c^m}, b \in v_k^{c^n}, ~and~ (a, b) \in L_{i,k}\}$. }

\subsection{Hub Participation ($\omega_h$)}
\label{sec:m3}

For many analysis, we are interested in knowing whether highly influential nodes within a community also interact across nodes in the other community. This can be translated to the \textit{participation of influential nodes within and across each participating community} for analysis. This is modeled by using the notion of  \textbf{hub}
\textbf{participation} within a community and their interaction across layers. In this paper, we have used degree centrality for this metric to connote higher influence. Ratio of participating hubs from each community and the edge fraction are multiplied to compute $\omega_h$. Formally,

For every $(u_i^m, u_k^n)$ $\in$ $L'_{i,k}$, where $u_i^m$ and $u_k^n$ are the meta nodes denoting the communities, $c_i^m$ and $c_k^n$ in the community bipartite graph, respectively, the weight,

\textit{\centerline{ $\omega_h$($u_i^m$, $u_k^n$) =  $\frac{|H_{i,k}^{m,n}|}{|H_i^m|}$ * $\frac{|x_{i,k}|}{|v_i^{c^m}|*|v_k^{c^n}|}$*$\frac{|H_{k,i}^{n,m}|}{|H_k^n|}$,}
}
\textit{\noindent where $x_{i,k}$ = $\bigcup~~\{(a, b): a \in v_i^{c^m}, b \in v_k^{c^n} ~and~ (a, b) \in L_{i,j}\}$; $H_i^m$ and $H_k^n$ are set of hubs in $c_i^m$ and $c_k^n$, respectively; $H_{i,k}^{m,n}$ is the set of hubs from $c_i^m$ that are connected to $c_k^n$; $H_{k,i}^{n,m}$ is the set of hubs from $c_k^n$ that are connected to $c_i^m$ }.


\subsection{Density and Edge Fraction ($\omega_d$)}
\label{sec:m2}

The intuition behind this metric is to bring participating community density which captures internal structure of a community. Clearly, \textit{higher the densities and larger the edge fraction, the stronger is the interaction (or coupling) between two meta nodes (or communities.)} Since each of these three components (each being a fraction) increases the strength of the inter-layer coupling, they are  multiplied to generate the weight of the meta edge.  The domain of this weight will be $(0,1]$. The weight computation formula is similar to the previous one and hence not shown. 

For every $(u_i^m, u_k^n)$ $\in$ $L'_{i,j}$, where $u_i^m$ and $u_k^n$ denote the communities, $c_i^m$ and $c_k^n$ in the community bipartite graph, respectively, the weight,

\textit{\centerline{$\omega_d$($u_i^m$, $u_k^n$) =  $\frac{2*|e_i^{c^m}|}{|v_i^{c^m}|*(|v_i^{c^m}|-1)}$ * $\frac{|x_{i,k}|}{|v_i^{c^m}|*|v_k^{c^n}|}$*$\frac{2*|e_k^{c^n}|}{|v_k^{c^n}|*(|v_k^{c^n}|-1)}$,}
}
\textit{\noindent where $x_{i,k}$ = $\bigcup~~ \{(a, b): a \in v_i^{c^m}, b \in v_k^{c^n}, ~and~ (a, b) \in L_{i,k}\}$}


Ideally, the alternatives for metrics should be independent of each other so they are useful for different analysis. Also, it is important that their computation be efficient (see Section~\ref{sec:cost-analysis}.) We believe that the three metrics proposed satisfy the above and our experiments have confirmed them although not shown in this paper due to page constraints.

\subsection{k-community Computation Efficiency}
\label{sec:cost-analysis}

For a given specification of a k-community, its computation has several cost components. Below, we summarize their individual complexity and cost.

\begin{enumerate}
    \item \textbf{Cost of generating 1-community}: For each layer (or a subset of needed layers) this can  be done in parallel bounding this \textbf{one-time cost} to the largest one (typically for a layer with maximum density.)
    
    \item \textbf{Cost of computing meta edge weights}: For the proposed analysis metrics, part of them, again, are \textbf{one-time costs} and are calculated independently on the results of 1-community. The  costs for $\omega_d$ and $\omega_h$ require a single pass of the communities  using their node/edge details generated by the community detection algorithm.

    \item \textbf{The recurring cost }(for each application of $\Theta$): This includes the cost of generating the bipartite graph, computing the weight of each meta edge of the community bipartite graph for a given $\omega$, and the MWxx algorithm cost. Only the edge fraction (or the maximum number of edges) and participating hubs need to be computed during each iteration. The cost of MWxx algorithm used in our experiments is \textit{almost the same as} the cost of computing MWM.  The bipartite graph is generated during the computation of weights for the meta edges. Luckily, in our community bipartite graph, the number of meta edges is \textbf{order of magnitude less} than the number of edges between layers. Also, the number of meta nodes is bound by the number of pairings in the previous iteration. MWM requires
    computing the weight of meta edges incident on a meta node. Extension of MWM to MWRM and MWMT requires at most a single pass of meta edges of the bipartite graph.
    
    
\end{enumerate}

\section{Mapping to k-Community Expressions}
\label{sec:application-and-analysis}

We introduce the data sets and analysis objectives to formulate the k-community specification along with the choice of weight metric to apply the algorithm. This will clarify the usage of the three weight metrics for MWxx algorithm. The same data sets are used for experimental analysis.
\noindent \textbf{DBLP data set~\cite{data/type2/DBLP}:} The DBLP data set captures information about published research papers in conference/journal, year of publication and the authors. Most readers are familiar with this data set.\\




\noindent\textbf{DBLP detailed Analysis Objectives}
	\begin{enumerate}[label={(A\arabic*)}]

    


\item \label{analysis:DBLPHe-PAu-wd-MWMT} For each conference, which are the \textit{most cohesive} group(s) of authors who \textit{publish frequently} (ties included)?

2-community: \textbf{P $\Theta_{P,Au}$ Au; $\omega_d$; $\Theta$ = MWMT}

    

    
    

\item \label{analysis:DBLPHe-PAuY-we-MWM} For the \textit{most popular unique} collaborators from each conference, which are the \textit{unique} 3-year period(s) when they were \textit{most active}? 

3-community: \textbf{P $\Theta_{P,Au}$ Au $\Theta_{Au,Y}$ Y; $\omega_e$; $\Theta$ = MWM}

\end{enumerate}

Based on the DBLP analysis requirements, three layers are modeled for the HeMLN\footnote{For both data sets, modeling of MLNs using analysis objectives as well as mapping to appropriate weight metric are beyond the scope of this paper. See ~\cite{msThesis/Komar19} for details.}. Layer \textit{Au} connects any two authors (nodes) who have published at least three research papers together. Layer \textit{P} connects research papers (nodes) that appear in the same conference. Layer \textit{Y} connects two year nodes if they belong to same pre-defined period. The inter-layer edges depict \textit{wrote-paper ($L_{Au,P}$), active-in-year ($L_{Au,Y}$) and published-in-year ($L_{P,Y}$)}. For this paper, we have chosen all papers that were published from 2001-2018 in top conferences. Six 3-year periods have been chosen: [2001-2003], [2004-2006], ..., [2016-2018]. \\

\noindent \textbf{IMDb data set~\cite{data/type2/IMDb}: }The IMDb data set captures movies, TV episodes, actor, directors and other related information, such as rating.\\

\noindent\textbf{IMDb Detailed Analysis Objectives}
	\begin{enumerate}[label={(A\arabic*)}, resume]



    

 

\item \label{analysis:IMdbHe-ad-we-MWRM} Find the actor and director similar-genre based group pairs such that \textit{overall actor-director collaborations are maximized}?
    
2-community: \textbf{A $\Theta_{A,D}$ D; $\omega_e$; $\Theta$ = MWRM}

\item \label{analysis:IMdbHe-ad-wh-MWPM} Based on similarity of genres, list the \textit{maximum number} of \textit{unique} actor and director groups whose \textit{majority of the most versatile members interact}?
    
2-community: \textbf{A $\Theta_{A,D}$ D; $\omega_h$; $\Theta$ = MWPM}


    

    

	

\item \label{analysis:IMdbHe-madm-we-MWMT}For the \textit{most popular unique} actor groups (including ties), from each movie rating class, find the \textit{unique} director groups with which they have \textit{maximum interaction} and who also make movies with similar ratings. 
	
Cyclic 3-community: \\ \textbf{M $\Theta_{M,A}$ A $\Theta_{A,D}$ D $\Theta_{D,M}$ M;  $\omega_e$; $\Theta$ = MWMT}

	\end{enumerate}

For addressing the IMDb analysis requirements, three layers for the IMDb data set are formed as follows. Layer \textit{A} and Layer \textit{D} connect actors and directors who act-in or direct \textit{similar genres frequently} (intra-layer edges), respectively. Layer \textit{M} connects movies within the same rating range. The inter-layer edges depict \textit{acts-in-a-movie ($L_{A,M}$), directs-movie ($L_{D,M}$) and directs-actor ($L_{A,D}$)}. There are multiple ways of quantifying the similarity of actors and directors based on movie genres they have worked in. A vector was generated with the number of movies for each genre he/she has acted-in/directed. In order to consider the similarity with respect to \textit{frequency of genres}, two actors/directors are connected if the Pearsons' Correlation between their corresponding genre vectors is at least 0.9\footnote{The choice of the coefficient is not arbitrary as it reflects relationship quality. The choice of this value can be based on how actors/directors are weighted against the genres. We have chosen 0.9 for connecting actors in their top genres.}. Moreover, 10 ranges are considered - [0-1), [1-2), ..., [9-10] for movie ratings.


\textbf{Choice of weight metric: } 
For the objectives specified in this paper, \textit{maximum interaction} and \textit{most popular} in ~\ref{analysis:DBLPHe-PAuY-we-MWM}, \ref{analysis:IMdbHe-ad-we-MWRM} and ~\ref{analysis:IMdbHe-madm-we-MWMT}, are interpreted as the number of edges between the participating communities. In contrast, interaction with cohesive groups as in~\ref{analysis:DBLPHe-PAu-wd-MWMT}, is interpreted to include community density as well. Versatility is mapped to participation of hub nodes in each group as in  ~\ref{analysis:IMdbHe-ad-wh-MWPM}.

\textbf{Choice of pairing algorithm:} Each pairing algorithm maximizes the overall weight based on a constraint. For \ref{analysis:DBLPHe-PAuY-we-MWM}, MWM is chosen due to the the unique pairing constraint. For \ref{analysis:DBLPHe-PAu-wd-MWMT} and \ref{analysis:IMdbHe-madm-we-MWMT}, the unique constraint is relaxed to only inlcude ties, thus MWMT is selected. For \ref{analysis:IMdbHe-ad-wh-MWPM}, the uniqueness criterion is combined with maximizing the number of pairs, thus we chose MWPM. In \ref{analysis:IMdbHe-ad-we-MWRM}, the uniqueness restriction is absent making MWRM the choice.

\textbf{Identifying the k-community:} 
\ref{analysis:DBLPHe-PAu-wd-MWMT}, \ref{analysis:IMdbHe-ad-we-MWRM} and \ref{analysis:IMdbHe-ad-wh-MWPM} compute a 2-community. ~\ref{analysis:DBLPHe-PAuY-we-MWM} requires a 3-community (for 3 layers) with an acyclic specification (using only 2 edges). 
~\ref{analysis:IMdbHe-madm-we-MWMT} uses the layer M twice for a  3-community and is also cyclic. Note that the analysis objectives have been chosen carefully to cover the weights and pairing algorithms discussed in the paper. The limitation on the number of analysis objectives is purely due to space constraints.


\section{Experimental Analysis}
\label{sec:experiments}
The choice of data sets and sizes used in this paper are primarily  for demonstrating the versatility of analysis using the k-community detection and its efficiency as well as drill-down  capability based on structure and semantics preservation. We are not trying to demonstrate scalability in this paper. Also, instead of presenting all communities, we have chosen to show a few important drill-down results to showcase the structure and semantics preservation of our approach.


\noindent \textbf{Experiment Setup:} For DBLP HeMLN, research papers published from 2001-2018 in VLDB, SIGMOD, KDD, ICDM, DaWaK and DASFAA were chosen. For IMDb HeMLN,  we extracted, for the top 500 actors, the movies they have worked in (7500+ movies with 4500+ directors). The actor set was repopulated with the co-actors from these movies, giving a total of 9000+ actors. For this set of actors, directors and movies, the HeMLN with 3 layers described in Section \ref{sec:application-and-analysis} was built.
Widely used Louvain method~\cite{DBLP:Louvain} is used to detect 1-communities. 
The k-community detection algorithm \ref{alg:k-community} was implemented in Python version 3.7.3 and was executed on a quad-core $8^{th}$ generation Intel i7 processor Windows 10 machine with 8 GB RAM. 


\subsection{Analysis Results}

\noindent\textbf{Individual Layer Statistics}:
Table \ref{table:IMDbHeMLNStats} shows the  layer-wise statistics for IMDb HeMLN. 63 Actor (A) and 61 Director (D) communities based on similar genres are generated. Out of the 10 ranges (communities) in the movie (M) layer, most of the movies were rated in the range [6-7), while least popular rating was [1-2). No movie had a rating in the range [0-1).  
\begin{table}[h!t]
\renewcommand{\arraystretch}{1}
\centering
    \begin{tabular}{|c|c|c|c|}
        \hline
      IMDb  & \textbf{Actor} & \textbf{Director} & \textbf{Movie} \\
        \hline
        \textbf{\#Nodes} & 9485 & 4510 & 7951 \\
        \hline
        \textbf{\#Edges} & 996,527 & 250,845 & 8,777,618  \\
        \hline
        \textbf{\#Communities (Size $>$1/all)} & 63/190 & 61/190 & 9/9 \\
        \hline
        \textbf{Avg. Community Size} & 148.5 & 73 & 883.4 \\
        \hline
        \end{tabular}
\caption{IMDb HeMLN Statistics}
\label{table:IMDbHeMLNStats}
    
\end{table}

\begin{table}[h!t]
\renewcommand{\arraystretch}{1}
\centering
    \begin{tabular}{|c|c|c|c|}
        \hline
      DBLP   & \textbf{Author} & \textbf{Paper} & \textbf{Year} \\
        \hline
        \textbf{\#Nodes} & 16,918 & 10,326 & 18 \\
        \hline
        \textbf{\#Edges} & 2,483 & 12,044,080 & 18 \\
        \hline
        \textbf{\#Communities (Size $>$1/all)} & 591/15528 & 6/6 & 6/6 \\
        \hline
        \textbf{Avg. Community Size} & 3.3 & 1721 & 3 \\
        \hline
        \end{tabular}
\caption{DBLP HeMLN Statistics}
\label{table:DBLPHeMLNStats}
    
\end{table}

Similarly, DBLP HeMLN statistics are shown in Table \ref{table:DBLPHeMLNStats}. 591 Author (Au) communities are generated based on co-authorship. 6 Paper (P) communities are formed by grouping papers published in same conference. KDD (2942) and DASFAA (583) have highest and least published papers, respectively. Out of 6 ranges of years (Y) selected, the maximum and minimum papers were published in 2016-2018 (1978) and 2001-2003 (1421), respectively.

\begin{table}[h]
\renewcommand{\arraystretch}{1}
\centering
    \begin{tabular}{|p{1.8cm}|p{1.25cm}|p{1.25cm}|p{1.25cm}|p{1.25cm}|}
        \hline
    \textbf{Expression}    & \textbf{MWM} & \textbf{MWMT} & \textbf{MWPM} & \textbf{MWRM} \\
        \hline


        
\textbf{A-D},~$\omega_e$ \textit{\#Comm::A(190), D(190)} & \textit{total}:83, \textit{partial}:0,  $\Sigma\omega$:9941 & \textit{total}:220, \textit{partial}:0,  $\Sigma\omega$:10127 & \textit{total}:95, \textit{partial}:0,  $\Sigma\omega$:\textit{1032} & \textit{total}:83, \textit{partial}:0,  $\Sigma\omega$:\textbf{11751} \\
        
        \hline
  \hline

\textbf{M-A-D-M},$\omega_e$ \textit{\#Comm::A(190), D(190), M(9)} & \textit{total}:2, \textit{partial}:7,  $\Sigma\omega$:6979 & \textit{total}:3, \textit{partial}:12,  $\Sigma\omega$:6984 & \textit{total}:0, \textit{partial}:9,  $\Sigma\omega$:\textit{6979} & \textit{total}:2, \textit{partial}:11,  $\Sigma\omega$:\textbf{11557} \\
        \hline
        \end{tabular}
\caption{Effect of Pairing Algorithms on same specification}
\label{table:analysis-statistics}
    
\end{table}

\textbf{Effect of Pairing Algorithm: 
} For an expression with a specified layer order of evaluation and weight metric, the results will vary based on choice of  algorithm for pairing communities from the community bipartite graph. This is illustrated in Table \ref{table:analysis-statistics} where we list the number of HeMLN communities (total+partial) and total  meta edge weights for the first round of matching ($\sum${$\omega$}) obtained  with 4 different pairing algorithms for  the same specification. 
It can be observed that  MWM generates the least number of pairs with maximum total weight. On the other hand, when the uniqueness condition is relaxed, the overall sum of weights is  improved by MWMT and MWRM.

\begin{figure}[htb]
   \centering
   \includegraphics[width=0.7\linewidth]{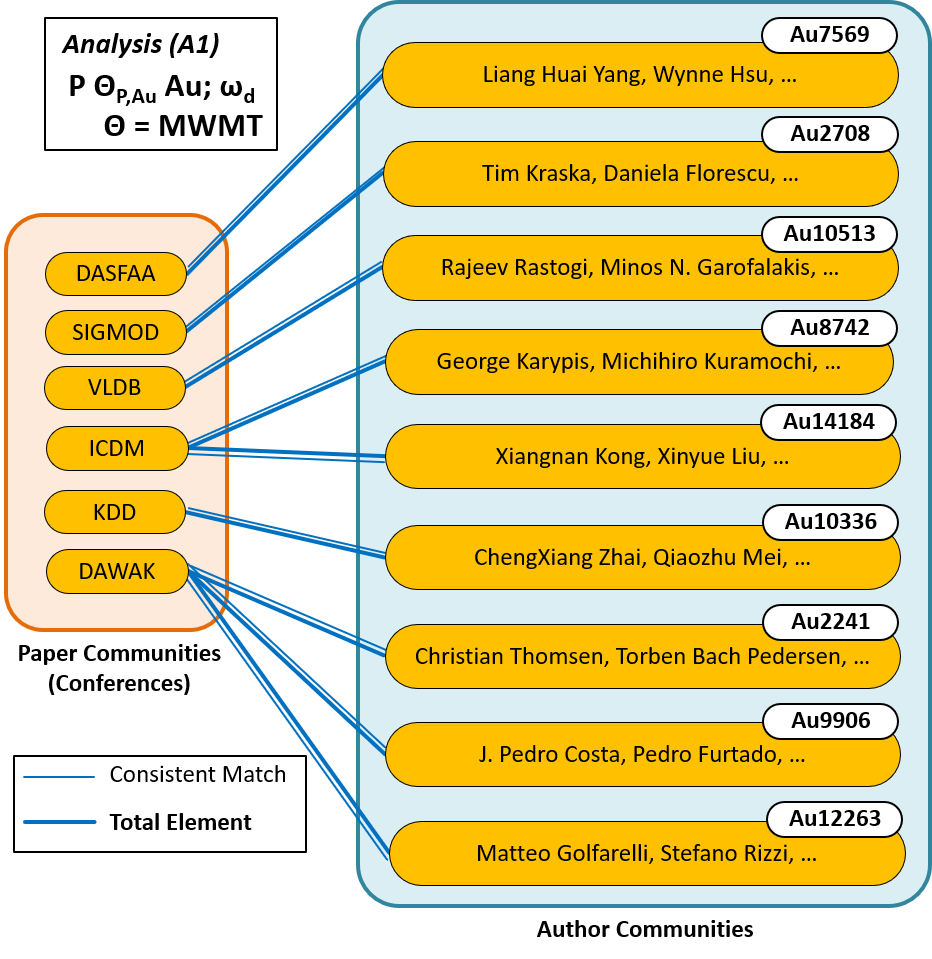}
   \caption{{\ref{analysis:DBLPHe-PAu-wd-MWMT} Result}: \textbf{9 Total Elements\protect\footnotemark}}
   \label{fig:DBLPHe-PAu-wd-MWMT}
\end{figure}

\footnotetext{Louvain numbers all communities from 1 and we only consider communities having \textit{at least two members} for this paper. The numbering used in the paper have layer name followed by the Louvain-generated community ID (e.g. A91, Au8742).}

\noindent\textbf{\ref{analysis:DBLPHe-PAu-wd-MWMT} Analysis: }
On applying MWMT on the CBG created with all Paper and Author communities, we obtained 9 total elements that correspond to the \textit{most cohesive co-authors who also publish frequently in each conference} (shown in Figure \ref{fig:DBLPHe-PAu-wd-MWMT} with list of few prominent authors.) ICDM and DaWaK have \textbf{multiple author communities} that are \textbf{equally important}. Prominent researchers like Tim Kraska and Daniela Florescu; Rajeev Rastogi and Minos N. Garofalakis, and;  George Karypis and Michihiro Kuramochi are members of the frequently publishing co-author group (in the last 18 years) for SIGMOD, VLDB and ICDM, respectively. These results can be validated from the facts that a) \textit{Tim Kraska} has been a recipient of \textbf{Best of SIGMOD Award (2008, 2016)}, b) \textit{Rajeev Rastogi}'s published papers in VLDB (in past 18 years) have received over 900 citations c) \textit{George Karypis} has been a recipient of \textbf{IEEE ICDM 10-Year Highest-Impact Paper Award (2010) and IEEE ICDM Research Contributions Award (2017)} \footnote{Intra-layer edge weights are not considered in this analysis. Hence, for an author (e.g., Jiawei Han) who has authored large \textit{number of papers}, his co-authors are distributed among different co-author communities due to lack of weight information and hence does not come out. This clearly demonstrates the need for weighted communities at the layer level to increase analysis space as has been shown with meta edge weights.}. 


 \begin{figure}[h]
   \centering
   \includegraphics[width=\linewidth]{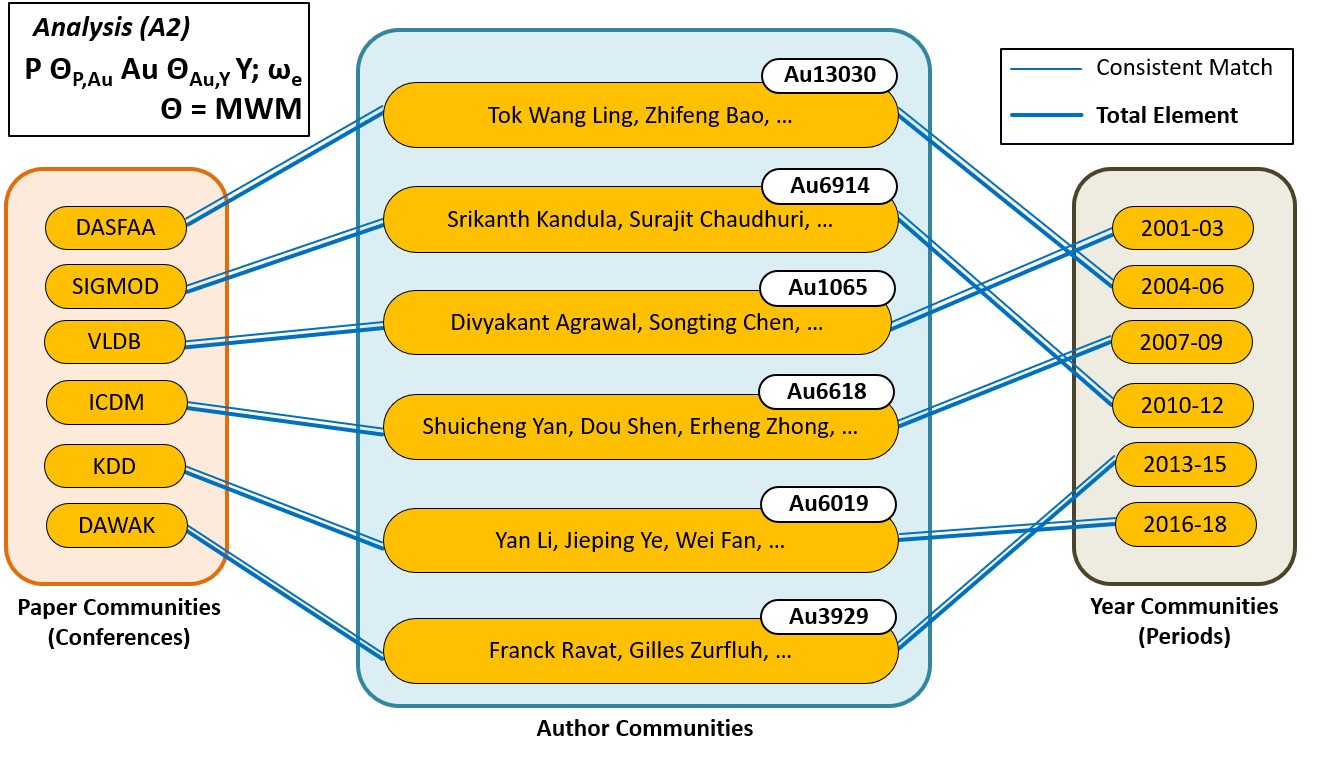}
   \caption{{\ref{analysis:DBLPHe-PAuY-we-MWM} Result}: \textbf{6 Total Elements}}
   \label{fig:A2-DBLP-PAuY-we-MWM}
\end{figure}

\noindent\textbf{ \ref{analysis:DBLPHe-PAuY-we-MWM} Analysis: } 
For the required \textit{acyclic 3-community} results, the \textit{most popular unique author groups} for \textit{each conference} are obtained by MWM (first composition). The matched 6 author communities are carried forward to find the \textit{disjoint year periods} in which they were \textit{most active} (second composition). 6 total elements are obtained (path shown by \textbf{\textcolor{blue}{bold blue lines}} in Figure \ref{fig:A2-DBLP-PAuY-we-MWM}.) 
Few prominent names have been shown in the Figure \ref{fig:A2-DBLP-PAuY-we-MWM} based on citation count (from Google Scholar profiles.) 
For example, for \textit{SIGMOD, VLDB and ICDM} the most popular researchers include \textbf{Srikanth Kandula (15188 citations), Divyakant Agrawal (23727 citations) and Shuicheng Yan (52294 citations)}, respectively who were active in different periods in the past 18 years.

 \begin{figure}[h]
   \centering
   \includegraphics[width=0.8\linewidth]{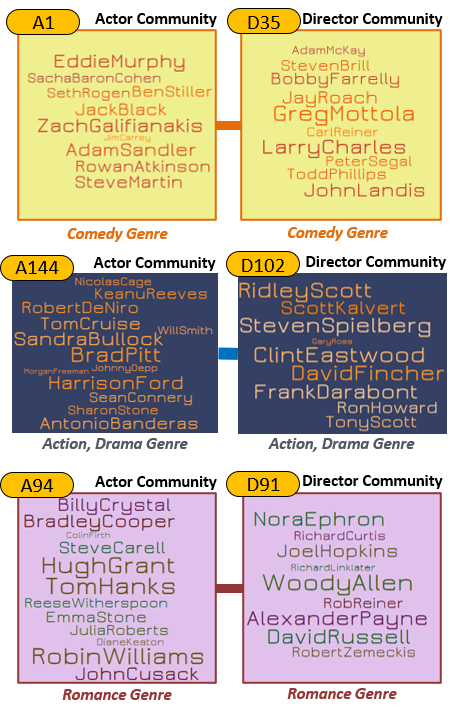}
   \caption{{Sample \ref{analysis:IMdbHe-ad-wh-MWPM} Result} for \textit{Comedy, Action/Drama, Romance} Genres}
   \label{fig:A4-IMDB-AD-wh-MWPM}
\end{figure}

\noindent\textbf{\ref{analysis:IMdbHe-ad-we-MWRM} Results: } 83 A-D (Actor-Director) similar genre-based \textit{overlapping} community pairs were obtained by MWRM, that maximised the \textit{overall number of actor-director interactions}. Due to the absence of uniqueness criterion, some actor communities were paired with multiple director communities, and vice-versa.

\noindent\textbf{\ref{analysis:IMdbHe-ad-wh-MWPM} Results: } MWPM \textit{maximizes the number of unique} A-D (Actor-Director) similar genre-based community pairs (29), where \textit{majority of most versatile members interact}. Intuitively, a group of directors that prominently makes movies in some genre (say, Drama, Comedy, Romance, ...) must pair up with the group of actors who primarily act in similar kind of movies. This can be validated from the few sample similar genre-based pairings  shown in Figure \ref{fig:A4-IMDB-AD-wh-MWPM} (drill down) , such as a) \textbf{Comedy} - Directors like \textbf{Bobby Farrelly, Todd Phillips, John Landis} etc. (from D35) pair up with actors like \textbf{Jim Carrey, Zach Galifianakis and Eddie Murphy} (from A1),  b) \textbf{Action/Drama} - Directors like \textbf{Clint Eastwood, Ridley Scott and Steven Spielberg} (from D102) pair up with Actors like \textbf{Brad Pitt, Tom Cruise and Will Smith} (from A144) and c) \textbf{Romance} -  Directors \textbf{Woody Allen, Tim Burton} etc. (from D91) pair up with the actors like \textbf{Diane Keaton, Emma Stone and Hugh Grant} (from A94). 



 \begin{figure}[h]
   \centering
   \includegraphics[width=\linewidth]{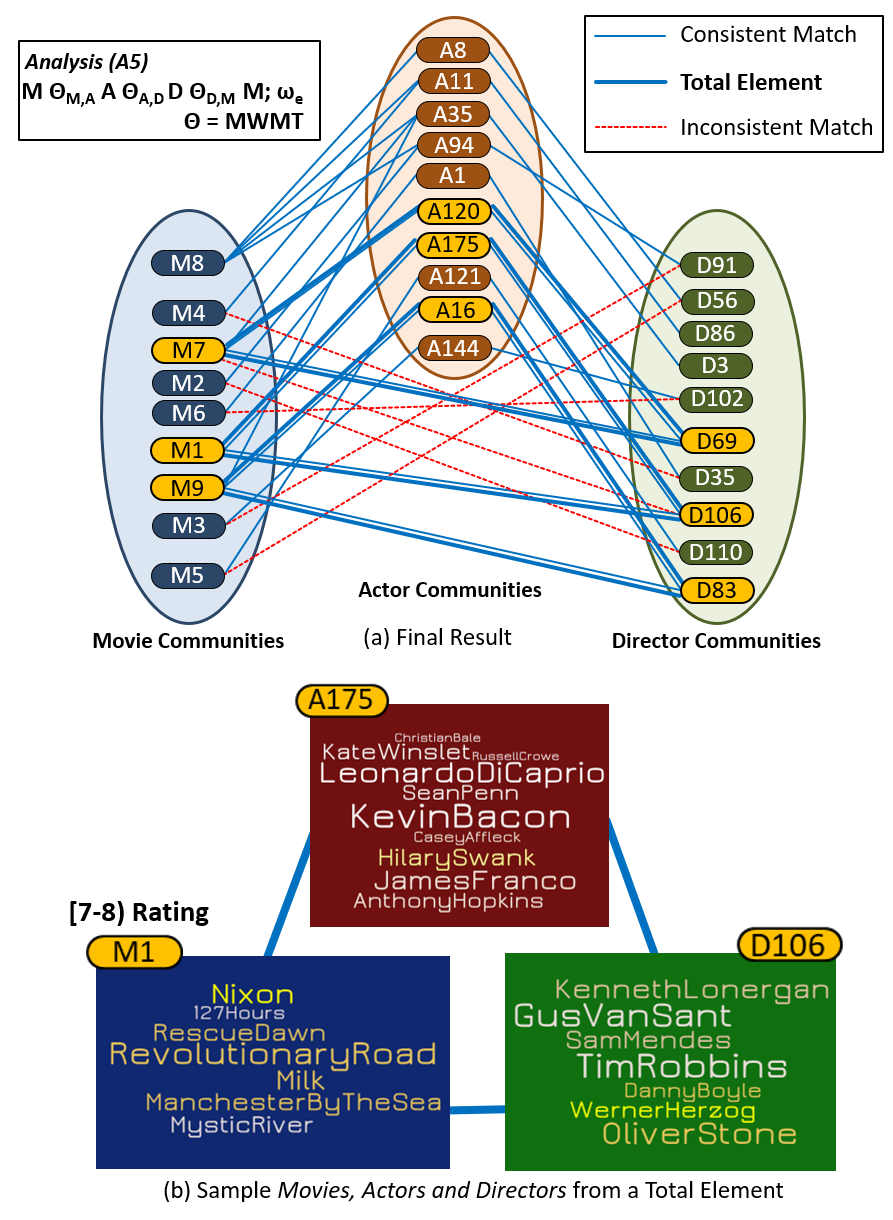}
   \caption{{\ref{analysis:IMdbHe-madm-we-MWMT} Result}: \textbf{3 Total, 12 Partial Elements}}

   \label{fig:A5-IMDB-MADM-we-MWMT}
\end{figure}

\noindent\textbf{\ref{analysis:IMdbHe-madm-we-MWMT} Results: }
Here, the \textit{most popular unique actor groups for each movie rating class are further coupled with directors}. These \textit{unique director groups are coupled again with movies to check whether the director groups also have similar ratings}. In every round for every pairing the ties are also included (MWMT). 
Results of each successive pairing (there are 3) are shown in  Figure  \ref{fig:A5-IMDB-MADM-we-MWMT} (a) using  the  same  color  notation. Coupling of movie and actor communities (first composition) results in 14 consistent matches. 
In the second composition with the director layer, using all director communities and the matched 10 actor communities, we got 10 consistent matches. The final composition to complete the cycle uses 10 director communities  and  9 movie communities as left and right sets of community bipartite graph,  respectively.

\textbf{Only 3 consistent matches are obtained to generate the 3 total elements  for the cyclic 3-community (\textcolor{blue}{bold blue triangle}.)} The total element M1-A175-D106-M1 
(sample members
shown in Figure \ref{fig:A5-IMDB-MADM-we-MWMT} (b)) \textit{groups together popular highly rated} (\textbf{Average Rating of [7-8)}) \textit{Drama} genre-based actors like \textbf{Leonardo DiCaprio, Sean Penn, Kate Winslet, Hilary Swank, Kevin Bacon, Anthony Hopkins, Russell Crowe, Christian Bale, James Franco and Casey Affleck} (from A175) with \textit{popular drama directors} like  \textbf{Danny Boyle, Sam Mendes, Werner Herzog, Gus Van Sant, Tim Robbins, Oliver Stone, Kenneth Lonergan}. This actor-director group is involved in few of the iconic award wining masterpieces like \textbf{Revolutionary Road, 127 Hours, Rescue Dawn, Milk, Mystic River, Nixon and Manchester By The Sea}. Most importantly, this genre-based group is also able to flesh out \textit{potential actor-actor or actor-director collaborations} like \textbf{DiCaprio-Swank-Mendes} and \textbf{Bacon-Hopkins-Stone}, who have not worked together yet.

It is interesting to see 6 inconsistent matches (\textcolor{red}{red broken lines}) between the communities  which  clearly indicate  that  all  couplings  are not  satisfied  by  these  pairs. This  results  in  12  partial elements which represent the similar genre-based actor and director groups but with \textit{different most popular movie rating classes}.

\textbf{The inconsistent matches also highlight the importance of mapping an analysis objective to a k-community specification for computation.} If a different order had been chosen (viz. director and actor layer as the base case), the result could have included the inconsistent matches. 



\subsection{Efficiency of Decoupled Approach}

\begin{figure}[ht]
   \centering
    \includegraphics[width=\linewidth]{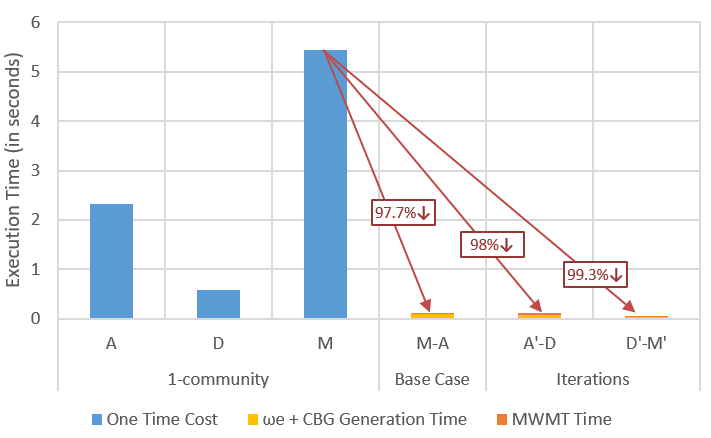}
   \caption{Performance Results for a 3-community using \ref{analysis:IMdbHe-madm-we-MWMT}}
   \label{fig:perf-madm}
\end{figure}

The goal of the decoupling approach was to preserve the structure as well as improve the efficiency of k-community detection using the divide and conquer approach. We illustrate that with the largest k-community we have computed which uses 3 compositions.
Figure \ref{fig:perf-madm} shows the execution time for the one-time and iterative costs discussed earlier for \ref{analysis:IMdbHe-madm-we-MWMT}. The difference in one-time 1-community cost for the 3 layers follow their density shown in Table~\ref{table:IMDbHeMLNStats}. We can also see how the iterative cost is insignificant as compared to the one time cost (by an order of magnitude.) Cost of each iteration includes creating the bipartite graph, computing $\omega_e$ for meta edges, and MWxx (in this case MWMT) cost. 
\textbf{The cost of all iterations together (0.27 sec) is more than \textit{an order of magnitude less than the largest one-time cost} (5.43 sec for Movie layer.)} We have used this case as this subsumes all other cases. 
The \textbf{additional incremental cost for computing a k-community is extremely small validating the efficiency of decoupled approach}. 


\section{Conclusions}
\label{sec:conclusions}

In this paper, we have provided a new structure and semantics preserving definition of a community for a HeMLN, two bipartite algorithms (MWRM and MWMT) suited for community detection, and an efficient ``decoupling-based" computation model. We have also demonstrated the ease with which drill-down of the results can be accomplished because of structure and semantics preservation.  \textit{Also, with  $\omega$ and MWxx as  customizable parameters, our approach supports a wide range of analysis objectives and  is extensible.}  We have compared our community definition with the traditional one using modularity to show their compatibility and closeness. Finally, we used the proposed approach for demonstrating its analysis versatility using both bipartite graph match algorithms and $\omega$ over the IMDb and DBLP data sets.

\ifCLASSOPTIONcompsoc
\else
\fi

\ifCLASSOPTIONcaptionsoff
  \newpage
\fi



%
\bibliographystyle{IEEEtran}

\bibliography{bibliography/santraResearch,bibliography/somu_research,bibliography/itlabPublications,bibliography/itlabTheses}

%

\end{document}